\documentclass[preprint,12pt]{elsarticle}

\usepackage{amssymb}
\usepackage{amsmath}

\usepackage{subcaption}
\usepackage{adjustbox}
\usepackage{multirow}
\usepackage[textwidth=3.2cm]{todonotes}

\usepackage[colorlinks=true, linkcolor=blue, citecolor=red, urlcolor=blue]{hyperref}
\usepackage{ifthen}

\newboolean{moreSections}
\setboolean{moreSections}{false} 
\usepackage{textgreek}

\journal{Journal of Systems Architecture}

\begin{document}

\begin{frontmatter}

\title{On the Power Saving in High-Speed Ethernet-based Networks for Supercomputers and Data Centers}

\author[uclm]{Miguel Sánchez de la Rosa}  
\author[uva]{Francisco J. Andújar}  
\author[uclm]{Jesus Escudero-Sahuquillo}  
\author[uclm]{José L. Sánchez}  
\author[uclm]{Francisco J. Alfaro-Cortés}  
\affiliation[uclm]{organization={Department of Computing Systems, Universidad de Castilla-La Mancha},
            addressline={Avda. de España},
            city={Albacete},
            postcode={02071},
            state={Castilla-La Mancha},
            country={Spain}}

\affiliation[uva]{organization={Departamento de Informática, Universidad de Valladolid},
            addressline={Plaza de Santa Cruz, 8},
            city={Valladolid},
            postcode={47002},
            state={Castilla y León},
            country={Spain}}

\begin{abstract}
The increase in computation and storage has led to a significant growth in the scale of systems powering applications and services, raising concerns about sustainability and operational costs. In this paper, we explore power-saving techniques in high-performance computing (HPC) and datacenter networks, and their relation with performance degradation. From this premise, we propose leveraging Energy Efficient Ethernet (EEE) protocol, with the flexibility to extend to conventional Ethernet or upcoming Ethernet-derived interconnect versions of BXI and Omnipath.

We analyze the PerfBound power-saving mechanism, identifying possible improvements and modeling it into a simulation framework. Through different experiments, we examine its impact on performance and determine the most appropriate interconnect. We also study traffic patterns generated by selected HPC and machine learning applications to evaluate the behavior of power-saving techniques.

From these experiments, we provide an analysis of how applications affect system and network energy consumption. Based on this, we disclose the weakness of dynamic power-down mechanisms and propose an approach that improves energy reduction with minimal or no performance penalty. 
To the best of our knowledge, this work presents the first thorough analysis of PerfBound and an enhancement to the technique, while also targeting emerging post-exascale networks.

\end{abstract}


\end{frontmatter}

\section{Introduction}
\label{sec:introduction}

Supercomputers and data centers are a critical infrastructure for IT applications and services.
The demand for increasing computing power and storage capacity from fields like Physics, Chemistry, Biology, or Medicine requires not only raw computing power, but also sophisticated and efficient supercomputers and data centers that provide the highest performance out of the available resources. The emergence of generative AI has produced a steep increase in computational demand, and it is expected to further expand and permeate into other fields and services. However, the high-performance computing demand comes at the cost of power consumption. Energy and environmental constraints must be considered when planning the infrastructure that powers the backbones of the mentioned applications and services.

Energy consumption in supercomputers and data centers is the sum of power drawn, among other elements, by computing and storage nodes, the interconnection network, and the cooling system. 
Considering the most powerful supercomputers, the Exascale barrier was first overrun by the Frontier supercomputer, which remained the first in the TOP500 ranking~\cite{top500} since June 2022, until El Capitan surpassed it in November 2024. \tablename~\ref{tab:top500top5} contains five of the most powerful supercomputers (in terms of computing power) with energy efficiency information supplied as per the last ranking. Energy efficiency is the ratio between the maximum performance and the average power consumption; the higher it is, the better. As we can see, the highest energy efficiency is achieved by the JUPITER booster machine, which is \#4 in the TOP500 ranking as of November 2025.

\begin{table}[htb!]
\centering
\resizebox{.75\textwidth}{!}{%
\begin{tabular}{ccccc}
\hline
\textbf{Rank} & \textbf{Name} & \textbf{\begin{tabular}[c]{@{}c@{}}Rmax\\ {[}PFlop/s{]}\end{tabular}} & \textbf{\begin{tabular}[c]{@{}c@{}}Power\\ {[}MW{]}\end{tabular}} & \textbf{\begin{tabular}[c]{@{}c@{}}Energy Efficiency \\ {[}GFlops/Watts{]}\end{tabular}} \\ \hline
1 & El Capitan & 1809 & 29.68 & 60,94 \\
2 & Frontier & 1353 & 24.61 & 54.98 \\
3 & Aurora & 1012 & 38.7 & 26.15 \\
4 & JUPITER Booster & 1000 & 15.8 & 63.32 \\
6 & HPC6 & 477.9 & 8.46 & 56.48 \\ \hline
\end{tabular}
}
\caption{Power consumption and energy efficiency in the first five ranks of TOP500.}
\label{tab:top500top5}
\end{table}

The operational costs for supercomputers are concerning because of power availability and sustainability reasons. Green500~\cite{green500} ranks TOP500 systems according to their power efficiency rather than the raw performance. If we examine the Green500 ranking for that date, we can see that the aforementioned systems are, in terms of power efficiency, in ranks \#23, \#34, \#90, \#14 and \#32, respectively. Of course, designing a system with high regard for power efficiency leads to lower operational costs, namely in electricity.

Processors and accelerators in modern supercomputers include mechanisms to reduce power under thermal or power constraints. Intel CPUs, for instance, introduced C-states~\cite{intelsandybridge} to enable advanced power management. These states trade performance and latency for lower power consumption and can be requested by the operating system. State transitions require microarchitectural operations and voltage changes, introducing latency, during which no code is executed. Moreover, even the deepest low-power states still consume energy. Overall, these mechanisms aim to make CPU energy consumption proportional to utilization. 

In this context, as computing nodes become more energy proportional, the interconnection network consumes more power, relatively~\cite{energyproportional}. 
The power contribution of the network can then increase, reaching up to nearly 40~\%~\cite{svidirovpower}.
Therefore, it is also necessary to design high-performance interconnection networks whose energy consumption is somewhat proportional to the traffic they carry. 

\begin{figure}[h!]
	\centering
	\begin{subfigure}[t]{0.49\textwidth}
		\centering
		\includegraphics[width=\textwidth]{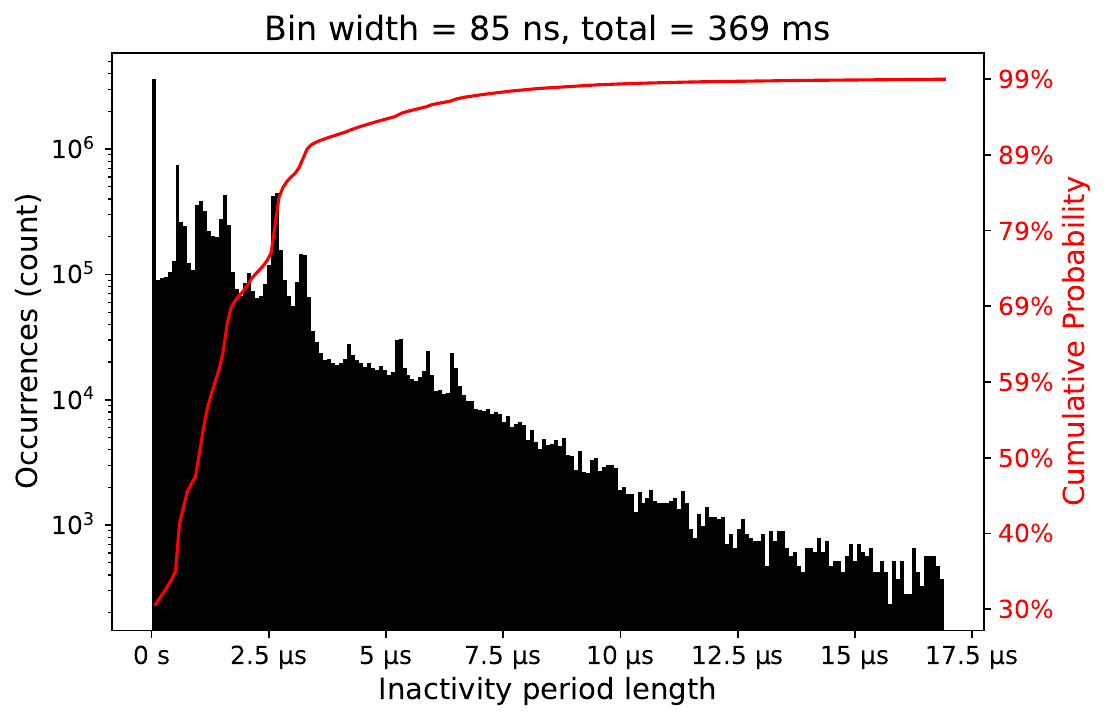}
		\caption{LAMMPS}
		\label{fig:inac-cdf-LAMMPS}
	\end{subfigure}
    \begin{subfigure}[t]{0.49\textwidth}
		\centering
		\includegraphics[width=\textwidth]{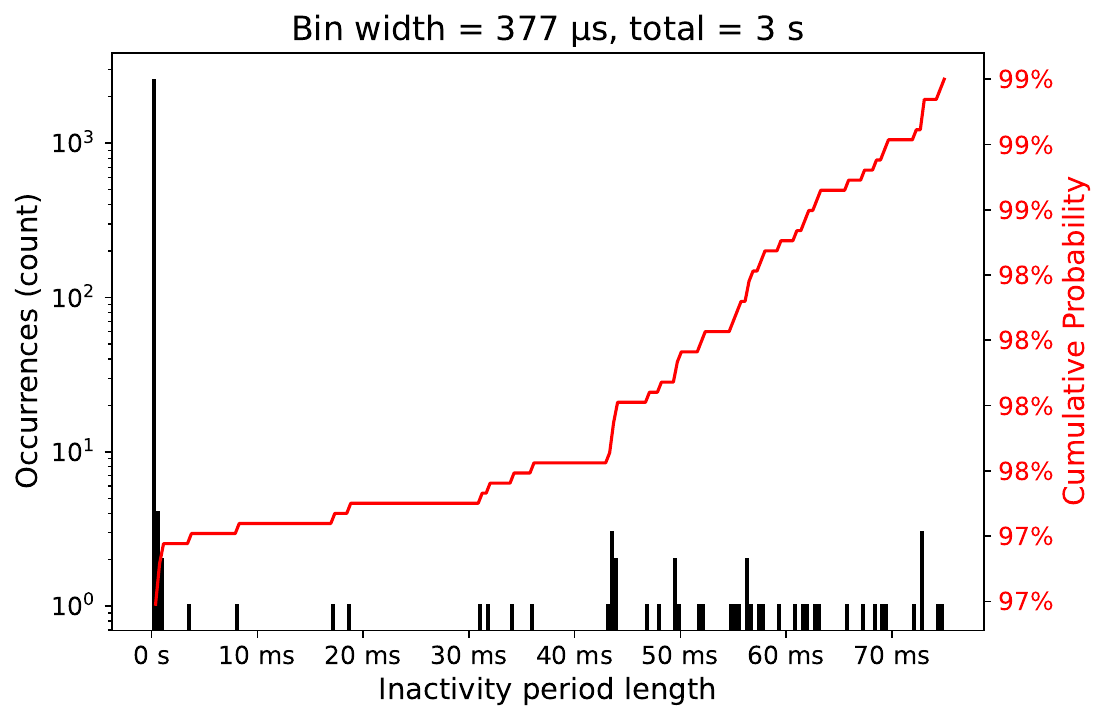}
		\caption{PATMOS}
		\label{fig:inac-cdf-PATMOS}
	\end{subfigure}
    
	\begin{subfigure}[t]{0.49\textwidth}
		\centering
		\includegraphics[width=\textwidth]{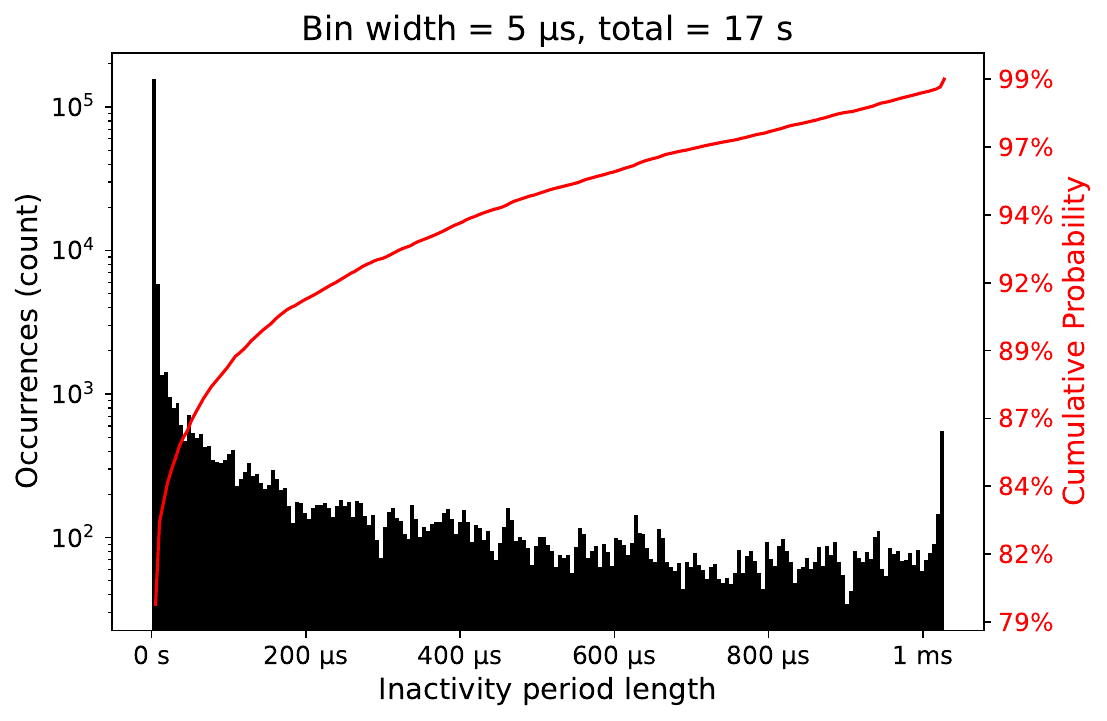}
		\caption{MLWF}
		\label{fig:inac-cdf-MLWF}
	\end{subfigure}
    \begin{subfigure}[t]{0.49\textwidth}
		\centering
		\includegraphics[width=\textwidth]{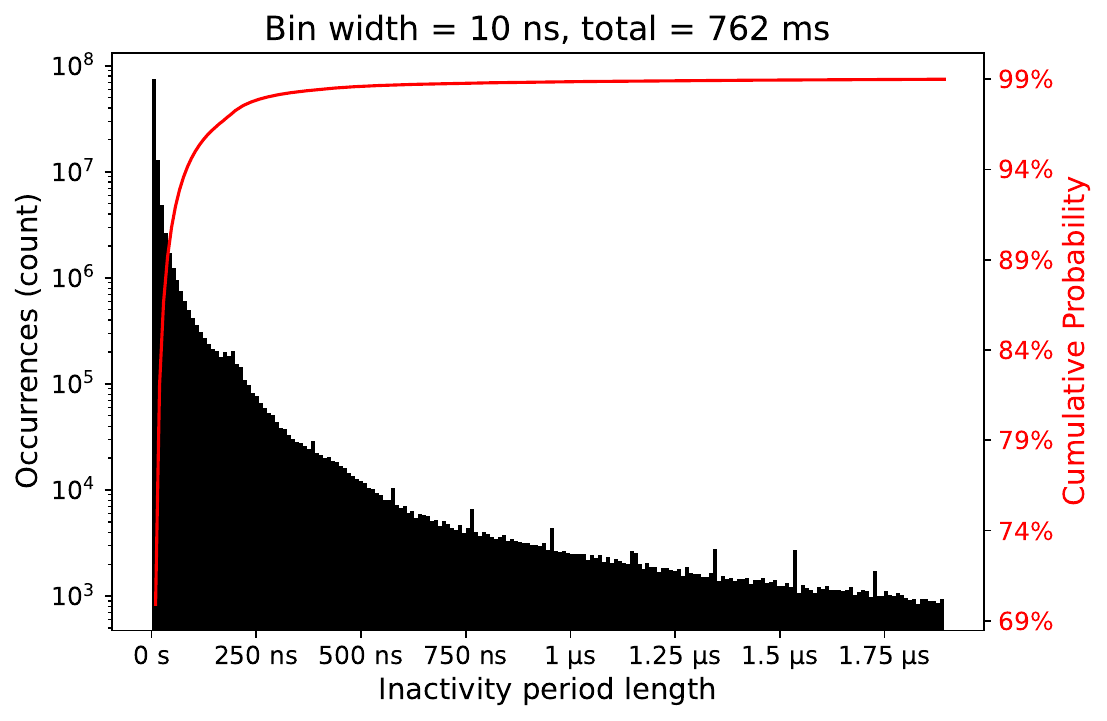}
		\caption{ALEXNET}
		\label{fig:inac-cdf-ALEXNET}
	\end{subfigure}
    
	\caption{Inactivity histogram for a port when executing different applications.}
    \label{fig:inac-cdf}
    
\end{figure}

Mainly, the opportunities for power savings come from periods of inactivity that would allow us to set ports to power-saving states. If we look at the network usage for several applications, we can see that network activity fluctuates during execution. Periods of low network usage are good opportunities for applying power-saving techniques. \figurename~\ref{fig:inac-cdf} shows the inactivity periods for a specific port on several applications when executed on our target simulated environment. The data is then used to form a histogram of 200 fixed-sized bins. Each bin $i$ stores the number of registered inactivity periods that last between $width_{bin} \times i$ and $width_{bin} \times (i+1)$ seconds. In these plots, we show values up to the 99\textsuperscript{th} percentile for visibility reasons. On the $x$ axis, we show the length of an inactivity period on that port. The left $y_1$ axis in black shows how many periods of inactivity are within the specified bin, and on the right, $y_2$ in red shows the cumulative distribution for every bin. In other words, the red line represents the percentage of the histogram's ``area'' below the bin. Regardless of the distribution of inactivity periods, any power-saving mechanism should take advantage of the longer periods to provide as much energy savings as possible. Moreover, we expect the performance penalty of the mechanism to be as low as possible or within a reasonable margin.

This paper explores the effects of network power management strategies on the performance of interconnection networks. First, we describe the background concepts in network energy reduction (see Section~\ref{sec:background}). Next, we describe the network power for large Ethernet-based interconnection networks used in supercomputers and data centers, and how it can be applied to BXIv3-based networks (see Section~\ref{sec:power-model}). We also show that more efficient interconnection networks are required to achieve energy proportionality. Afterwards, we describe the power model for BXIv3 that we have designed and the proposed energy-saving strategies.
This model has been implemented in a powerful simulation framework, allowing large-scale network simulations.
Furthermore, we have performed a set of simulation experiments (see Section~\ref{sec:evaluation}) that show that, with the proposed power model strategies, the network power can be lowered without gravely affecting the overhead on execution time or latency. In particular, we managed to save more than 5~\% of energy on the network by focusing on links only, while still outperforming other previous proposals in terms of performance overhead. We manage to improve the PerfBound~\cite{perfbound} technique, even to the point of reverting energy increase into actual savings. In our opinion, this is a promising start, given that our proposal and assumptions could be portable to future interconnection network technologies, such as BXIv3 and the upcoming (Formerly Intel's) Omnipath\footnote{
\href{https://www.theregister.com/2025/06/09/omnipath_is_back/}{https://www.theregister.com/2025/06/09/omnipath\_is\_back/}}~\cite{omnipath}.

\ifthenelse{\boolean{moreSections}}{
\subsubsection{Problem Statement}
}{}

In summary, the main contributions of this paper are the following:
\begin{itemize}
    \item {Study of the power consumption and efficiency in HPC interconnection networks.}
    \item {Exploration of proposals to reduce link power draw in said networks, tailored to HPC and datacenter workloads.}
    \item {Study and implementation discussion of PerfBound~\cite{perfbound}, focused on the BXIv3 interconnect.}
    \item {A proposal (PerfBoundCorrect) to mitigate the oversights and implementation details of PerfBound.}
\end{itemize}

\section{Background}
\label{sec:background}

\subsection{Energy consumption in high-speed interconnection networks}
Network performance is affected by several network design issues, such as the topology, routing, flow control, and operating frequency, among others. Thus, designing an efficient and cost-effective network is crucial to guarantee the performance of the entire supercomputer and datacenter, as well as a balance between this performance and the investment and operational costs. Regarding HPC network technologies, a quick look into the TOP500 list shows us that Gigabit Ethernet, in various revisions, is present in many systems (34.2~\%), only second to Infiniband~\cite{ibta2025spec} at 55.4~\%. 
Such is the momentum of Ethernet in these environments that technologies like iWarp~\cite{iwarp} and RoCE (RDMA over Converged Ethernet)~\cite{roceiwarp} are working towards providing Ethernet with RDMA, present in interconnects like Infiniband.

Beyond the substantial capital expenditure (CAPEX), operational costs (OPEX), and energy constraints are critical in HPC systems. Major OPEX components include power supply, cooling, and maintenance. Energy-saving efforts focus on reducing power consumption, potentially at the cost of performance, while cooling demands scale with heat dissipation (a direct consequence of power). At the HPC scale, both CAPEX and OPEX are enormous, and electricity costs can be decisive. Thus, reducing power consumption is a key objective, particularly given the sparse network communication in many scientific applications. In fact,  InfiniBand, the most pervasive interconnect in HPC, has power-saving mechanisms~\cite{dickov_software-managed_2014}. The ubiquity of Ethernet in other commercial areas has also caused several proposals to appear for Ethernet as well~\cite{gunaratne_reducing_2008,hays_activeidle_2008}. The IEEE Task Force assembled to provide the technology with a common standard, resulting in the standard IEEE 802.3az, also known as Energy Efficient Ethernet (EEE)~\cite{specifications_ieee_2009}.

Several design and operation strategies have been applied to reduce energy consumption in interconnection networks.
Specifically, network energy consumption can be reduced by designing networks with high-radix switches. By increasing the switch radix, fewer routers are required. As switches aggregate more components, this also reduces the number of hops~\cite{kim_microarchitecture_2005}. Consequently, the number of links in the network is reduced when the hop count is lower. Routing algorithms can also be relevant to energy consumption if additional links provide several paths between two nodes. Of course, these links will consume energy when unused. 

Operational costs (OPEX) in HPC systems increase with the number of nodes and network components. As with processors, network energy consumption can be reduced during periods of inactivity, which are dictated by application communication patterns. Scientific simulations typically minimize inter-node communication, resulting in sparse network activity. In contrast, data center workloads such as deep learning, distributed storage, cloud computing, and search engines exhibit network loads driven by user requests and strict latency requirements~\cite{andujar_extending_2023}. Reducing power in these multi-tenant environments is cumbersome due to certain unpredictability in network activity.

\subsection{Initial attempts for reducing link power consumption}

Device power consumption comprises static and dynamic components, with dynamic power depending on clock frequency, supply voltage, and transistor technology. Since power scales roughly with the cube of frequency, substantial savings are possible. Dynamic Voltage and Frequency Scaling (DVFS) adapts voltage and frequency to utilization but requires additional hardware, such as voltage regulators, increasing complexity and conflicting with the goal of simple interconnection networks~\cite{kim_adaptive_2002, shang_dynamic_2003}.

As network technologies often support multiple data rates, Adaptive Link Rate (ALR) was proposed to reduce energy by lowering link speeds during low utilization without changing voltage or frequency. However, rate transitions incur latency due to MAC handshakes, PHY reconfiguration, and resynchronization~\cite{gunaratne_reducing_2008}, limiting adoption and potentially creating bottlenecks in HPC workloads.

\subsection{Energy-Efficient Ethernet and Low Power Idle}\label{sec:background:EEE-lpi}
As ALR proved inefficient, Low Power Idle (LPI) was proposed, standardized, and used by EEE~\cite{specifications_ieee_2009,christensen_ieee_2010}. LPI specifies two port power states,
\emph{Wake} for ports ready to transmit, and \emph{Sleep}, a power-saving state, for inactive ones. This standard also specifies the basic signaling behavior, but the transition policies and implementation details are left for vendors.
The LPI key point is that higher data rates are more power-efficient in terms of Joules per bit transmitted. Therefore, it would be more suitable than ``downgrading'' links as in ALR. The transition to \emph{Sleep} consists of shutting down power-hungry PHY components so that transitions are shorter than those of ALR. However, links still draw power because they must synchronize periodically. 

\begin{figure}[!htb]
    \centering
    \includegraphics[width=.75\textwidth]{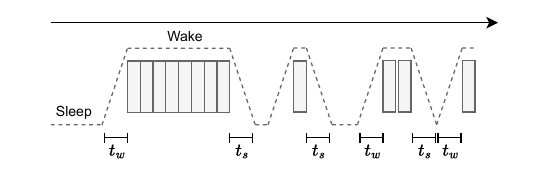}
    \caption{Port switching between \emph{Wake} and \emph{Sleep} states due to packet transmissions.}
    \label{fig:lpi}
\end{figure}

Figure~\ref{fig:lpi} illustrates the interaction between packet transmissions and LPI state transitions. The parameters ${t}_{w}$ and ${t}_{s}$ denote the time required to transition from \emph{Sleep} to \emph{Wake} and vice versa.
When packets are routed through a link in the \emph{Sleep} state, the transmission can only begin after $t_w$, when the transition ends. Long message bursts of contiguous packets compensate for the transition overhead (one transition for the whole message), whereas short and sporadic packets may experience more frequent transition delays. These communications happen often in HPC workloads. Power consumption during state transitions is considered to be as much as in the \emph{Wake} state. This damages energy saving twice, due to both the transition overhead and the high power draw.

Table~\ref{tab:lpi-timings} shows the $t_w$ and $t_s$ transition times defined by IEEE 802.3az for Ethernet rates up to 10GBase-T, along with the transmission time of a 1500-byte frame for comparison. While link speeds increase by an order of magnitude, transition times do not decrease proportionally. Still, LPI is suitable for long bursts followed by periods of inactivity comparatively longer than transmission time. Those inactive times will provide the most opportunities for power saving.

\begin{table}[h!]
\centering
\resizebox{.75\textwidth}{!}{%
\begin{tabular}{cccc}
\hline
\textbf{Protocol} & \textbf{${t}_{w}$} & \textbf{${t}_{s}$} & \textbf{Frame time (1500B)} \\ \hline
100Base-Tx (100 Mb/s) & 30 \textmu s & 100 \textmu s & 120 \textmu s \\
1000Base-T (1 Gb/s) & 16 \textmu s & 182 \textmu s & 12 \textmu s \\
10GBase-T (10 Gb/s) & 4.48 \textmu s & 2.88 \textmu s & 1.2 \textmu s \\ \hline
\end{tabular}%
}
\caption{Link transition times for different Ethernet versions (extracted from~\cite{specifications_ieee_2009}).}
\label{tab:lpi-timings}
\end{table}

\begin{figure}[!htb]
    \centering
    \includegraphics[width=\textwidth]{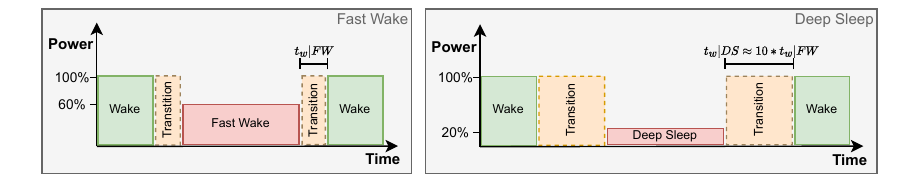}
    \caption{Visual representation of \emph{Deep Sleep} and \emph{Fast Wake} power states.}
    \label{fig:fw-ds}
\end{figure}

Subsequent IEEE amendments refined the original LPI model to mitigate wake-up latency at higher link speeds. In particular, IEEE 802.3bj, targeting 40~Gb/s and faster Ethernet links, introduced the \emph{Fast Wake} state and renamed the original low-power mode to \emph{Deep Sleep}~\cite{noauthor_options_2012}. \emph{Fast Wake} keeps more components of the PHY on, reducing $t_w$ to a tenth compared to Deep Sleep, at the cost of reduced power savings (20–40~\% compared to up to 90~\% in \emph{Deep Sleep})~\cite{saravanan_powerperformance_2013}. Figure~\ref{fig:fw-ds} conceptually illustrates the difference between these low-power states.

These multiple power levels provide the basis for energy-aware link management strategies in Ethernet networks. In HPC environments, several proposals leverage \emph{Fast Wake} to limit the latency impact of LPI~\cite{reviriego_performance_2009,ANDUJAR2023126}. However, transitioning to low-power states instantly after a packet transmission may cause links to spend more time in state transitions than in active data transmission, reducing the potential energy savings.

IEEE specifications beyond 40GBase-T do not explicitly report LPI transition times, which depend on the PHY implementation and medium. For higher-speed links, we estimate wake and sleep transition durations using a power-law extrapolation based on measured PHY generations. Denoting the link rate as $R$, the wake-up time is modeled as $t_w \propto R^{b}$, yielding $\log(t_w) = b\,\log(R) + c$. A least-squares fit in log--log space results in a scaling exponent of $b = -0.45$ for wake transitions and $b = -0.77$ for sleep transitions.

Table~\ref{tab:lpi-estimated} summarizes the measured LPI transition times up to 10GBase-T and the extrapolated values for higher Ethernet rates. These estimates represent indicative trends rather than standard-compliant parameters.

\begin{table}[htb!]
\centering
\begin{tabular}{ccc}
\hline
\textbf{Ethernet rate} & $\boldsymbol{t_w}$ (\textmu s) & $\boldsymbol{t_s}$ (\textmu s) \\ \hline
100Base-TX & 30 (measured) & 100 (measured) \\
1000Base-T & 16 (measured) & 182 (measured) \\
10GBase-T & 4.48 (measured) & 2.88 (measured) \\ \hline
40G Ethernet & 2.4 (estimated) & 1.2 (estimated) \\
100G Ethernet & 1.6 (estimated) & 0.7 (estimated) \\
400G Ethernet & 0.8 (estimated) & 0.25 (estimated) \\ \hline
\end{tabular}
\caption{Measured and extrapolated LPI wake ($t_w$) and sleep ($t_s$) transition times as a function of Ethernet bandwidth.}
\label{tab:lpi-estimated}
\end{table}

\subsection{Proposals to complement EEE}
\label{sec:background:EEE-extensions}

The main basis behind power management in interconnection networks is that link power consumption does not depend on usage. However, link usage is not constant because transmissions only occur when necessary. Moreover, the usage of any particular link is determined by several factors. For example, the traffic pattern, the topology, and the routing algorithm can have different effects on the usage of every link during the execution of an application. Deciding on when to switch to and from a power-saving state is important because the more power savings, the more overhead packets can suffer.

The simplest power-saving strategy is to switch a port to a power-saving state as soon as a transmission is complete. When two connected ports have nothing to send, they can transition at the same time. However, switching to power-saving states for every transmission entails transitioning to the  \emph{Wake} state before the next transmission as well. Even during the most intense communication periods, traffic is not sent at once, and link usage is recurrent and intermittent.

\ifthenelse{\boolean{moreSections}}{
\subsubsection{Power-down Threshold}
}{}
The overhead for turning a port on is greatly compensated if the following transmission is comparatively higher. However, short and sparse packets increase transmission overhead, similarly.
If the port is left in  \emph{Wake} state for a moderate period after transmission, there is no overhead to the next packet transmitted within that time. Power-down Threshold (PDT)~\cite{saravanan_powerperformance_2013} was proposed so that every port stays operational for an additional time after transmission, at the cost of potential power saving. 

The technique can be implemented with a timer on each port. The port turns down when the timer expires after transmitting a packet. If any packet arrives at the port, either from the link or via arbitration or fragmentation, the timer is canceled. It is set again right after transmission, when there are no more packets to send.

\begin{figure}[!htb]
    \centering
    \vspace*{-.5cm}
    \includegraphics[width=.8\textwidth]{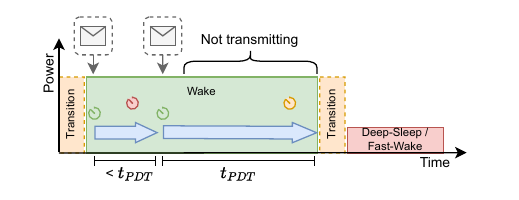}
    \vspace*{-.5cm}
    \caption{Diagram of LPI using PDT.}
    \label{fig:lpi-pdt}
\end{figure}

Figure~\ref{fig:lpi-pdt} shows link power behavior under PDT. Blue arrows indicate the timer duration, and the timer icon’s color reflects its state changes. Green means that it is set,  red is for reset, and salmon is for expired. During the \emph{Wake} state, packets present at the port can be sent normally.  After sending a packet, the port waits for a certain period (${t}_{PDT}$) to transition to the \emph{Sleep} state. If a packet arrives within ${t}_{PDT}$ after sending a packet, the timer is canceled and set again when the packet is transmitted. 

This way, frequently used ports will be in the \emph{Wake} state most of the time, but packets will not suffer from latency because ports are ready to send them. Meanwhile, for inactivity periods that are on the order of seconds, the overhead on latency for state transitioning is irrelevant. 
The issue, as we have mentioned, is the intermittent nature of network activity during communication. When the ${t}_{PDT}$ is on the same order of magnitude as the inactivity periods, ports may often receive packets after powering down, or even while transitioning.

\ifthenelse{\boolean{moreSections}}{
\subsubsection{PerfBound}
}{}

The main drawback of PDT is selecting an appropriate timer value and generalizing it for all ports. The former depends on the application and would require prior execution for inactivity profiling. This would be useful if the inactivity periods of the ports could be estimated beforehand. However, they depend on multiple factors, such as task mapping and the number of processors on the nodes, which affect the traffic pattern. In a production environment, the values would have to be fine-tuned for an application. The latter issue of PDT is that inactivity periods differ between all ports. This can be caused by network topology or task scheduling. The topology determines the average and maximum hop count for packets between a source and destination, both by the layout of network elements and the routing algorithm. In fact, the higher the hop count, the larger the overhead due to port transitions.

Due to these limitations, ports should establish their own ${t}_{PDT}$ value, based on previous network activity. To solve this issue, Saravanan proposed PerfBound~\cite{perfbound}. The basis of PerfBound is forming a histogram on each port with the inactivity periods so that the port can calculate its particular ${t}_{PDT}$ at any given moment in a way that degradation is bound. PerfBound also requires every port to store (or calculate) the distance in hops of any packet, given a destination.
In this way, we can quantify the possible degradation of a packet due to the powering up of multiple ports along its route.

To calculate the next ${t}_{PDT}$ value, the algorithm first determines the number of packets that can be delayed due to power-up transitions within a given degradation limit. For example, if we want to maintain a performance degradation restriction of up to 1~\% in a period of duration $X$, we can delay up to $N = 0.01*X/{t}_{w}$ packets, where ${t}_{w}$ is the transition time to \emph{Wake}. We use the average distance to destinations described earlier as a corrective factor $l$ for each port. By doing so, we are less restrictive on ports with a higher mean distance. In a generalized manner, $l = (bound\%) \times \sum_{i=1}^{n} \frac{p_i}{h_i}$, where $n$ is the number of different hop counts, $p_i$ is the proportion of packets that travel $h_i$ hops and $h_i$ is the number of hops for the $i$-th group of packets.

For example, if 70~\% of packets sent by a port are 6 hops away from the destination, and the distance for the remaining 30~\% is 6 hops, we would calculate that factor for a 1~\% degradation threshold as in Eq.~\ref{localbound}:

\begin{equation}
l = 0.7 \times \frac{0.01}{4} + 0.3 \times \frac{0.01}{6} = 0.01 \times \left( \frac{0.6}{4} + \frac{0.4}{6} \right) \approx 0.0023
\label{localbound}
\end{equation}

To apply this factor, the number of packets we can delay will be $N = l * X /{t}_{w}$. Once we have obtained $N$, we search on the histogram, from the highest to the lowest bin, and accumulate the count of each bin. We are looking for the leftmost bin that accumulates at most $N$. We will set the ${t}_{PDT}$ to the mean of that bin.

\subsection {BXIv3 network technology}

BXIv3~\cite{bxi3eviden} is the network technology being developed by Eviden for HPC systems and data centers. The main objective for Eviden is to develop an interconnect that can power post-exascale systems.
The existing versions of BXI, namely V1.2~\cite{bxihoti} and V2~\cite{bxi2atos}, are already deployed in several TOP500 supercomputers, including CEA-HF and Tera-1000-2~\cite{top500}. Choosing Ethernet as a basis means that there is a grounded work on standards that can speed up the development of the technology. Some examples of this set of standards are the 802.1Qbb or priority flow control (PFC)~\cite{cclsrdmad} for lossless per-queue link-level flow control, or the possibility to use variable-sized packets and jumbo frames.

BXIv3 network interfaces (NICs) support a link bandwidth of $400$ Gbps (expected to be $800$ Gbps in the near future), with a target deployment scale of up to 100,000 nodes, following the modular supercomputing architecture (MSA) specification~\cite{Suarez19MSA}. BXIv3 supports different network topologies, such as Fat-trees and Megafly/Dragonfly+~\cite{flajslik_megafly_2018,dfly+}, which permit the desired levels of interconnection scalability. BXIv3 switches assume a Combined Input-Output Queued (CIOQ) switch architecture, similar to other Ethernet-based options. 
Regarding power consumption, we are not aware of any proposal tailored to BXIv3\footnote{Further details on the BXIv3 architecture can be obtained from one of the papers describing the BXIv3 architecture in the RED-SEA project~\cite{Gomez24redsea}.}. We believe that, given the high consideration for power consumption, it will be a cornerstone for future interconnection technologies.

\section{Power consumption model}
\label{sec:power-model}

This section describes the interconnection network power model tailored to the BXIv3 architecture we implemented in our network simulator.
This model is based on the Energy Efficient Ethernet (EEE) standard (see Section~\ref{sec:background:EEE-lpi}) and the extensions proposed to expand its functionality (see Section~\ref{sec:background:EEE-extensions}), such as LPI, PDT, and \emph{PerfBound}.
We have also designed some improvements to the \emph{PerfBound} strategy, so that EEE can be better applied to Ethernet-based interconnection networks when used to run HPC applications.

For BXIv3-based interconnection networks, we assume a combined input-output queued (CIOQ) switch architecture. Specifically, BXIv3-based switches define buffers at input and output ports, so it is possible to grant switch crossing requests from input buffers to non-operative ports, i.e, to output ports in the \emph{Sleep} state. This way, packets will be stored on their respective output queues when ports power up, i.e., when returning to the \emph{Wake} state. 

We have opted to implement and experiment with the EEE standard because BXIv3 and further switch generations are Ethernet-based\footnote{\href{https://www.nextplatform.com/2024/12/10/eviden-mainstreams-bxi-interconnect-thanks-to-ultra-ethernet-and-ai-boom/}{https://tinyurl.com/a484fjpj}}.
Furthermore, Eviden, the French company devising BXI, is one of the founder members of the Ultra Ethernet Consortium (UEC), where Eviden has planned a Roadmap for BXIv4 and BXIv5 generations in this decade.
Since the UEC explicitly aims to scale relatively flat Ethernet networks to support up to $1$ million endpoints within the next several years, the power management will be a cornerstone to keep proportional energy consumption in these networks.

In the following sections, we describe the EEE-based power model details regarding link power states, transition logic using PDT, implementation of \emph{PerfBound}, and the improvements made to adapt these techniques to BXIv3 networks when running HPC applications.

\subsection{Link power-levels and transition logic}

The link power levels are as described in the EEE standard, with three states: \emph{Wake}, \emph{Fast Wake}, and \emph{Deep Sleep}. As the EEE only specifies the components to shut down and the corresponding timings to transit between them, it is also up to the vendors and network architects to decide when and how to switch states.

Our proposal for deciding on port state transition revolves around the state of the output queues. This way, ports that have sent packets will notify their remote ports to power down, synchronizing their transition.

\begin{figure}[h!]
    \centering
    \includegraphics[width=.85\textwidth]{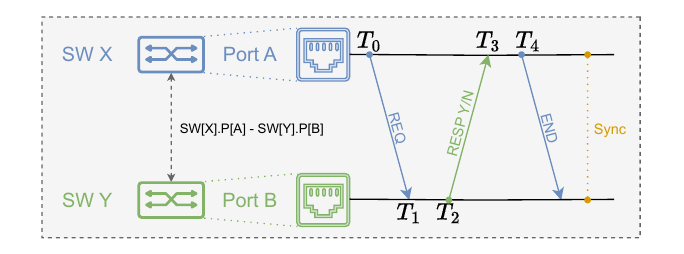}
    \caption{Port state synchronization between two switch ports.}
    \label{fig:port-sync}
\end{figure}

\figurename~\ref{fig:port-sync} shows the request-response mechanism that two ports from different devices, connected by a link, use to change states at the same time. At $T_0$, port A requests synchronization to either power up or down. Port B receives the request at $T_1$, and responds at $T_2$ with the answer. Port A has not only the response, but can also calculate the Round-Trip-Time (RTT) to establish the start of the transition, marked in the figure at `Sync'. Port A sends a message to signal the end of synchronization so that Port B can also calculate the RTT.

The criterion to accept or refuse to transition mainly depends on the port buffer occupancy. When a port is requested to power up, the response is always positive and the port switches to the \emph{Wake} state. Upon receiving a power-down request, a port responds with a rejection if its output buffer is empty, or with an ongoing transmission that the remote port may not be aware of. Otherwise, it responds affirmatively, and both ports synchronize to start the transition to the \emph{Sleep} state. 
This message exchange provides an insignificant overload: a 64-byte message encoding these messages requires 1.28 ns at 400 Gbps, which results in a delay below 5 ns in the case of a negative response to power down before the consequent packet transmission.

An output port in the \emph{Sleep} state may request to power up when it receives a packet from an input port via switch arbitration. In fact, it is the only way to trigger it because the remote input port at the next switch is also powered down. A control message is sent over the link to ensure the remote port transitions simultaneously, and the packet flowing over the link connecting these two switches is restored.
By contrast, if an output port is transitioning to power down and receives a packet via arbitration, it must start powering up as soon as it ends its transition down; otherwise, the synchronization between the two sides of the link becomes more complex. 

The PDT timer at a given switch port should always be set when a link is transmitting and expire only after all data has been sent from that port. In other words, expiration would mean that a port has already transferred all of its data. Therefore, when a port receives a packet in its output buffer, it resets the PDT timer to start counting after the packet is fully transmitted. The other port cancels the timer upon reception. This way, the last port transmitting is the one that ultimately requests to power down.

\subsection{PerfBound implementation details}
\label{sec:hist}

While PerfBound provides grounded methods to trade power saving for performance, some implementation and evaluation aspects are set under the authors' discretion.

For example, the histogram formed of inactivity period lengths is the key piece of the mechanism. However, even if clearing the histogram (or \textit{refreshing}, as the authors mention) is pondered, its consequences are only theoretically mentioned. There may be several possibilities for managing the histogram to constrain its storage size and maintain a bias towards recent values. Deciding on which information to remove from the histogram, based on time or size factors, can lead to different results, but it is necessary as storage is finite. The maintenance parameters of the histograms are shown in \tablename~\ref{tab:hist-params}. 

\begin{table}[htb!]
\centering
\begin{tabular}{@{}lll@{}}
\hline
\textbf{Parameter} & \textbf{Description} & \textbf{Encodes} \\ \hline
$n_{vals}$ & \# of stored values & Counter \\
${n}_{bins}$ & \# of populated histogram bins & Counter \\
$r_{bin}$ & Bin/element range (time unit) & Time \\
$w_{counter}$ & Bin counter width & Size (bytes) \\
$s_{elem}$ & Element size & Size (bytes) \\
${TTL}_{hist}$ & Histogram maximum time-to-live & Time \\
$h_{maxN}$ & Maximum records on histogram & Counter \\
$h_{maxval}$ & Maximum value to store in the histogram & Time \\ \hline
\end{tabular}
\caption{Histogram maintenance parameters}
\label{tab:hist-params}
\end{table}

The behavior upon collecting an inactivity time with value $t_{inac}$ (up to $h_{maxval}$) depends on the histogram management strategy. If the criterion to clean the histogram is to clear the whole structure, we only have to find the bin index for $t_{inac}$ and store the bin counter. $t_{inac}$ falls into bin ${inac}_{i}/r_{bin}$, where $r_{bin}$ represents a time unit, and each bin indexes a counter of $w_{counter}$ bytes. A map-like structure is ideal for this implementation, only storing counters for populated bins.

When the timer to wipe the histogram expires after ${TTL}_{hist}$ or after storing $h_{maxN}$ elements, the map is cleared. If we want to update the histogram in an informed manner, for example, by getting rid of the oldest values, we must track individual values and construct the histogram using them. We would implement this, for example, with a circular buffer of $h_{maxN}$ elements. Using that list, we can calculate the bin indices for each element and increment bin counters. The space requirements of these histogram implementations will be further described in section~\ref{sec:reqs}. In any of those cases, we can also set a ${haxval}_{hist}$ to prevent great outliers and reduce the total number of bins in the histogram.

Another important aspect of the algorithm is using average hop distances from switches and nodes to destination nodes. The experiments exploring the original work set a constant hop count for the network. However, this is non-realistic and may be affected by the traffic patterns of different applications. Devices should have their own record of hop distances, based on the destinations that can be reached according to their routing algorithm.
Device-to-destination distance can either be calculated theoretically (offline) using the routing algorithm or dynamically obtained with a traceroute-like all-to-all operation. For the second option, every source node sends this kind of message to each possible destination, increasing the TTL until it reaches the real hop count. Any intermediary device that sets TTL to 0 will then record the highest TTL value that subsequent messages from the same source to the same destination have. The highest TTL value recorded by a device will contain the exact hop count to the destination.

Another important aspect of PerfBound is that, while it provides a sensible approximation to a useful $t_{PDT}$ value, the final execution time overhead is unknown beforehand. For example, if we have an application that completes its execution in 100 seconds and we set PerfBound to only allow 1~\% overhead, the maximum possible overhead should be 1~second. During the execution, we will have multiple calculations for $t_{PDT}$ values on every port, and we will necessarily miss some predictions, degrading performance. 
The target overhead can be reached or surpassed during the application execution, but we have no way to notice (we assume we do not know the execution time beforehand). More importantly, the overhead is purely additive. Therefore, even if we could track the overhead during execution, we could not overcompensate for the threshold having been surpassed.  

The recurrent calculation of a PDT timer expiration value for every port is also concerning. More so for switches, which would need to do so every time all output queues for any port are emptied. We believe that this processing should be carried out regularly, but only if there was network activity on a port since the last calculation.

\subsection{Additional requirements}\label{sec:reqs}
PDT requires one timer and one register to store the configured ${t}_{PDT}$ in each port. The time resolution and maximum possible value are determining factors. For instance, a 32-bit unsigned register with up to nanosecond precision can encode ${2}^{32} * \frac{1s}{10^9ns} \approx 4.295$ seconds. \tablename~\ref{tab:maxpdt} shows the maximum time that could be encoded with several register widths and time resolutions.

\begin{table}[htb!]
\centering

\resizebox{.75\textwidth}{!}{%
\begin{tabular}{llllll}
\hline
\textbf{} & \multicolumn{1}{c}{\textbf{}} & \multicolumn{4}{c}{\textbf{Resolution}} \\
 & \textbf{} & \textbf{1 ns} & \textbf{10 ns} & \textbf{100 ns} & \textbf{1 \textmu s } \\ \hline
\multirow{5}{*}{\rotatebox{90}{\textbf{Width}}} 
 & \textbf{16 bits} &  65.54 \textmu s &  655.36 \textmu s   &   6.554 ms &  66 ms   \\
 & \textbf{32 bits} &  4.295 s  &  43 s  &  429 s  &  $4 \times 10^{3}$ s  \\
 & \textbf{48 bits} &  $2.81 \times 10^{5}$ s  &  $3 \times 10^{6}$ s  &  $28 \times 10^{6}$ s  &  $2.81 \times 10^{8}$ s  \\
 & \textbf{64 bits} &  $1.8 \times 10^{10}$ s  &  $1.84 \times 10^{11}$ s  &  $2 \times 10^{12}$ s  &  $1.8 \times 10^{13}$ s  \\
 & \textbf{128 bits} &  $3.4 \times 10^{29}$ s  &  $3 \times 10^{30}$ s  &  $3.4 \times 10^{31}$ s  &  $3.4 \times 10^{32}$ s  \\ \hline
\end{tabular}
}
\caption{Maximum time values to encode with several time resolutions on multiple register widths.}
\label{tab:maxpdt}
\end{table}

For PerfBound, each port requires some memory to form the histogram. For an estimation of the histogram memory footprint, we must look again at \tablename~\ref{tab:hist-params}. Mainly, the maximum histogram size depends on its maintenance strategy and the width for the bin counters ($w_{counter}$). The inactivity periods will be determined by the traffic pattern of the application, which will generate a variable number of histogram bins. If we only store the counters for every histogram bin, the histogram will take up $w_{counter} \times {n}_{bins}$~bytes, with ${n}_{bins} \leq h_{maxN}$. For a histogram of up to $h_{maxN}$ elements stored in a circular buffer, the extra memory requirement is $h_{maxN} \times s_{elem}$. In other words, up to $ h_{maxN} * (w_{counter} + s_{elem})$ bytes of memory if ${n}_{bins} = h_{maxN}$. Steep memory requirement increases beyond the range of MiBs are not expected if the histogram is cleared within a sensible period, like one second, and values are below the range of millions, with the original authors pointing out 20k elements as an example.

\subsection{Improvements to PerfBound: PerfBoundCorrect}

Even if PerfBound is a grounded basis for predicting a reasonable ${t}_{PDT}$ value, it presents mainly two drawbacks. The first one is that collecting previous values to calculate a new ${t}_{PDT}$ is ultimately a prediction for the next one. Prediction is widely used in decision-making on computers, from branching predictors to unattended maintenance at low-impact intervals.
In several of these scenarios, missing predictions often have the same performance overhead as not using the predictor; however, for PerfBound, the performance impact of every missed prediction is quantifiable. 

Every time we turn on a link (after $t_{inac}$ seconds of inactivity), it means that our last prediction missed because the last $t_{PDT}$ value is lower than $t_{inac}$. In other words, $t_{PDT}$ should have been ${t_{inac}}\over{t_{PDT}}$ ($>1$) times longer for the prediction to hit.
By measuring the frequency and magnitude of misses on recent predictions, we aim to enhance PerfBound with a correction factor that reduces future miss rate and thus, the performance overhead. 
We call this technique \emph{PerfBoundCorrect}. The only new parameter for this enhancement is the history length.

As this technique is built upon PerfBound, the same space requirements and histogram management strategies apply. Our proposal employs the output of the PerfBound technique and previous predictions to calculate a new $t_{PDT}$ value. The sequence is: collect inactivity periods, form the history/histogram, calculate the PerfBound prediction, and then calculate the PerfBoundCorrect factor. The enhancements to power efficiency come from applying the corrective factor, as the histogram is managed similarly.

Memory-wise, we have to store the miss history of the previous $n_R$ predictions, and the ratios that are above 1.

The miss history can be encoded in a shift register, and a FIFO queue of up to $n_R$ floating-point numbers can be used for the ratios. Every time a packet arrives at a port, if the PDT is still set (the port is up), we count this as a hit, or a miss otherwise. 
When a port starts its transition to power up, the ratio (${t_{inac}}/{t_{PDT}}$) is calculated and stored in a FIFO queue of length $n_R$ at the same time as the shift register is updated. If we remove a miss from the register by shifting, we also have to pop the oldest element from the queue. This way, we keep as many elements on the queue as misses encoded at the shift register. We only track missed predictions because we want to exclusively increase $t_{PDT}$ values using those ratios above 1. The proportion between the miss count and the shift register length returns the hit rate of the last $n_R$ predictions, and the queue of ratios can be used to calculate a correcting factor for the output of PerfBound. 

To apply the correcting factor to the PerfBound calculation, we have opted for the geometric mean of all ratios times the miss rate for the last $T_{PDT}$ value. The geometric mean was chosen because it is more tolerant of outliers than the arithmetic mean. The corrective factor ($cf$) applied to the output of PerfBound is as in Eq.~\ref{eq:pbcfactor}: 

\begin{equation}
cf = miss \% \times \left(\prod _{i=1}^{n}x_{i}\right)^{\frac {1}{n}}= miss \% \times {\sqrt[{n}]{r_{1}r_{2}\cdots r_{n}}}
\label{eq:pbcfactor}
\end{equation}

To prevent too small $cf$ values and result in unreasonably high (i.e., second range or above) $t_{PDT}$ values, we set a maximum limit $h_{maxval}$ within a sensible order of magnitude. The corrective factor $cf_b$ will be $min(cf,h_{maxval})$. The impact of great outliers in ratios that come from long inactivity periods is minimized by the miss rate: the lower the miss rate, the less these values affect the correcting factor. The influence of these outliers in the corrective factor diminishes as the network activity increases over time, and we reduce the miss rate.
Giving this leeway to the PerfBound algorithm may seem harmful to power saving, as links would spend more time operational after their transmissions end. However, it also has the potential to reduce the overhead caused by very reduced ${t}_{PDT}$ values. This is our proposal, and the results are non-trivial, as we will see in Section~\ref{sec:evaluation}.

\section{Evaluation}
\label{sec:evaluation}

In this section, we evaluate the power management mechanisms previously proposed, which we have modeled in our OMNeT++-based SAURON simulator~\cite{sauron}. First, we describe the experiment configurations and the network traffic workloads. Next, we include and analyze the experiment results for these workloads.

\subsection{Experiments configuration}

The SAURON simulator follows an event-driven approach and models the network devices (i.e., network interfaces and switches) at the packet level.
We have assumed the network parameters shown in \tablename~\ref{tab:exp-params}.

\begin{table}[htb!]
\centering
\begin{tabular}{ll}
\hline
\textbf{Parameter} & \textbf{Value} \\ \hline
Arbitration & Aging (crossbar), Round-Robin (output port) \\
Link speed & 400 Gbps \\
MTU & 9600 Bytes \\
Network topology & Megafly (see Section~\ref{sec:evaluation:topology}) \\
Node count & 4160 \\
PFC GO limit & 3200 Bytes (6.51 \% occupancy) \\
PFC STOP limit & 37900 Bytes (77.11 \% occupancy) \\
Queue size & 49152 Bytes \\
Switch architecture & CIOQ \\
Switch count & 1040 \\
Switch radix & 16 \\
Switch speed-up & 2.44 \\ \hline
\end{tabular}
\caption{System configuration parameters for the experiments.}
\label{tab:exp-params}
\end{table}

Regarding the network traffic, our simulator employs the open-source VEF Traces framework~\cite{VEF-traces, Andujar16JSC,andujar_extending_2023}, which allows us to record MPI-based traffic in the VEF traces and use the TraceLib library to inject those messages into the SAURON simulator.
The VEF traces have been collected in our own cluster CELLIA~\cite{cellia}, mainly composed of multiprocessor nodes. As we describe later, we have selected traces from different applications, such as HPC and Deep Learning.
The TraceLib library also provides information about the computing activity at system nodes, which helps us model their power consumption. Specifically, this power is a direct function of usage, with a minimum power draw at idle. Thus, the dynamic power consumption of a node ($P$) is mainly determined by its maximum and minimum power consumption, and the percentage of utilization based on computing activity (e.g., CPU or accelerator activity): $P = P_{min} + (P_{max}-P_{min})~\times  \%~utilization$.

Regarding the network topology, the SAURON simulator uses the TopGen~\cite{topgen} library and its file generator to model large-scale networks with multiple topologies. Specifically, we have selected the Megafly topology for these experiments. As we discuss in the next subsection, Megafly is more power-efficient compared to other alternatives.

\subsection{Discussion on the Topology selection}
\label{sec:evaluation:topology}

For the evaluation experiments, we assume a Megafly~\cite{flajslik_megafly_2018} topology (a.k.a., Dragonfly+~\cite{dfly+}) composed of 4160 computing nodes. The network is arranged in 65 groups of 64 nodes, each group employing 16 switches with a radix of 16 (1040 switches in total). The inter-group connectivity is as described in the original Megafly publication, with one global link containing every pair of groups. 
We have selected this topology for the experiments due to its scalability, flexibility, and growing popularity in HPC environments~\cite{mflyadapt}. 

Specifically, Megafly topologies can exploit the relative locality of nodes placed in the same group for fast efficient communication in disjoint network partitions that exchange their messages through high-bandwidth global channels. This node arrangement can help perform partial calculations in the same-group nodes and use the global links for reductions. AI workloads, and most recently, Large Language Models (LLMs), can also benefit from this topology, as several kinds of parallelization employ different broadcast domains. Compared to the Dragonfly, the main advantages are the use of low-radix switches and the inherent absence of deadlocks.
The routing algorithm modeled in the simulator is deterministic and minimal. When a packet has its source and destination in the same group, the algorithm behaves like D-mod-k~\cite{PGFT}. However, when the destination is in another group, the packet travels to the spine switch that will lead to the destination group through a global link. When the packet arrives at the destination group, the downward path is the same as in D-mod-k, with only one possible route.

Regarding power consumption, Megafly is a power-efficient alternative compared to other options such as 3D-Tori or RLFTs.
Table~\ref{tab:topo-power-all} shows a comparison in power consumption of the different devices of an HPC system (i.e., Switch, Node, and Links) for a Megafly, 3D-Torus, and RLFT networks using a similar number of node counts.

\begin{table}[!htb]
\centering
\resizebox{\textwidth}{!}{%
\begin{tabular}{c|ccccccc}
\hline
\textbf{} & \multirow{2}{*}{\textbf{Element}} & \textbf{Power/Unit} & \multirow{2}{*}{\textbf{Count}} & \multicolumn{3}{c}{\textbf{Power (all)}} & \textbf{} \\
\textbf{} &  & \textbf{Min} &  & \multicolumn{2}{c}{\textbf{Min}} & \multicolumn{2}{c}{\textbf{Max}} \\ \hline
\multirow{3}{*}{\textbf{Megafly}} & \textbf{Switch} & 250 W & 1040 & 260 KW & (6.36 \%) & 260 KW & (4.52 \%) \\
 & \textbf{Node} & 800 W - 1200 W & 4160 & 3.328 MW & (81.42 \%) & 4.992 MW & (86.80 \%) \\
 & \textbf{Link} & 24 W & 20800 & 499.2 KW & (12.21 \%) & 499.2 KW & (8.68 \%) \\ \cline{2-8} 
\textbf{} &  &  &  & 4.087 MW &  & 5.751 MW &  \\ \hline \hline
\multirow{3}{*}{\textbf{Torus}} & \textbf{Switch} & 250 W & 1000 & 250 KW & (6.07 \%) & 250 KW & (4.37 \%) \\
 & \textbf{Node} & 800 W - 1200 W & 4000 & 3.2 MW & (77.63 \%) & 4.800 MW & (83.89 \%) \\
 & \textbf{Link} & 24 W & 28000 & 672 KW & (16.3 \%) & 672 KW & (11.74 \%) \\ \cline{2-8} 
\textbf{} &  &  &  & 4.122 MW &  & 5.722 MW &  \\ \hline \hline
\multirow{3}{*}{\textbf{RLFT}} & \textbf{Switch} & 250 W & 845 & 211.250 KW & (4.85 \%) & 211.250 KW & (3.45 \%) \\
 & \textbf{Node} & 800 W - 1200 W & 4394 & 3.515 MW & (80.64 \%) & 5.273 MW & (86.20 \%) \\
 & \textbf{Link} & 24 W & 26364 & 632.736 KW & (14.52 \%) & 632.736 KW & (10.34 \%) \\ \cline{2-8} 
 &  &  &  & 4.359 MW &  & 6.117 MW &  \\ \hline
\end{tabular}%
}
\caption{Power consumption break-down in several networks.}
\label{tab:topo-power-all}
\end{table}

These numbers are based on information on modern server processors and expected network power draw provided by Eviden. The link power consumption is left constant before considering the power-saving features. The total link count is the number of links on every switch multiplied by the radix, plus the number of NICs. The system is considered idle when the nodes are not performing any computation and thus consume their minimum power (Min). Switch power only comprises the device power draw with no links connected.

On the contrary, under full load (Max), the system consumes approximately 3 MW for all the network topologies. Note that the network's power consumption is the sum of switch and link power. For instance, the network power for the Megafly is around 260~KW for switches and 499.2~KW for links, for a total of 759.2~KW. In proportion, the network accounts for 18.57~\% of power when the system is idling, and up to 13.2~\% under full load when all nodes are computing.
Compared to RLFTs and 3D-tori in this example, the Megafly achieves a lower network power consumption (the sum of switches and links), both at full load and when idling.

\subsection{Power saving parameters, network traffic and performance metrics}

In \tablename~\ref{tab:netovertotal}, we define the power parameters for each power-saving state, such as the power reduction and the time to transition from the power-saving state to the active state ($t_w$), and its counterpart ($t_s$), derived from the latest EEE specifications.
The table also shows the percentage of power that the system devotes to the network upon applying those values to our selected scenario. The calculi for the power contribution of different elements are similar to the one by~\cite{ANDUJAR2023126}. Every percentage shown in the table accounts for the contribution to the system energy. The \textbf{Network} column includes the combined power consumption of switches and links. The upper limit for power consumption on links is the difference between the \textbf{Wake} and Fast Wake/Deep Sleep contribution to power under full load.

\begin{table}[!htb]
\centering
\resizebox{\textwidth}{!}{%
\begin{tabular}{@{}ccccccccc@{}}
\hline
\multirow{2}{*}{\textbf{Port power state}} & \multicolumn{2}{c}{\multirow{2}{*}{\textbf{Power}}} & \multirow{2}{*}{\textbf{$t_w$}} & \multirow{2}{*}{\textbf{$t_s$}} & \multicolumn{2}{c}{\textbf{Links}} & \multicolumn{2}{c}{\textbf{Network}} \\
 & \multicolumn{2}{c}{} &  &  & \textbf{Idle} & \textbf{Full load} & \textbf{Idle} & \textbf{Full load} \\ \hline
\textbf{Wake} & 24 W &  & - & - & 12.21 \% & 8.68 \% & 18.58 \% & 13.20 \% \\
\textbf{Fast Wake} & 9.6 W & -60 \% & 375 ns & 200 ns & 4.89 \% & 3.66 \% & 20.04 \% & 13.93 \% \\
\textbf{Deep Sleep} & 2.4 W & -90 \% & 4.48 us & 2 us & 1.22 \% & 0.92 \% & 20.87 \% & 14.32 \% \\ \hline
\end{tabular}
}
\caption{Network contribution to total system power consumption across two scenarios.}
\label{tab:netovertotal}
\end{table}

For the evaluation, we have selected VEF traces of two HPC applications (LAMMPS and PATMOS) and two distributed Deep Learning training sessions (MLWF~\cite{MLWF} and AlexNet~\cite{ALEXNET}) to study the network behavior in both contexts. For each application, we will first describe the application behavior and the potential for energy saving. First, we will compare how much energy we save with different techniques in comparison to not using any power-saving technique.
Non-energy metrics include the degradation that is introduced by power-saving strategies. Mainly, we want to focus on the increase in execution time and latency. Specifically, the latency metric we are exploring is the packet latency, which measures the time from the packet generation to its arrival at the destination.

In the tests where a fixed PDT is employed, we have tested 9 possible $t_{PDT}$ values, from 0 to one second. For PerfBound and PerfBoundCorrect tests, we have tested 3 degradation thresholds (1, 2, and 5~\%) on each of the histogram types we discussed in section~\ref{sec:hist}.

\subsection{Evaluation for the LAMMPS application}

The LAMMPS molecular dynamics uses the network to exploit its parallel architecture and keeps the model distributed among nodes. The simulation space is divided into subdomains assigned to different processors, each of which manages \emph{own atoms} and \emph{ghost atoms}—copies of neighboring atoms needed to calculate interactions.

To maintain synchronization across subdomains, LAMMPS relies heavily on MPI communication, which generates traffic in two main forms. Point-to-point (P2P) messages are used to transfer data between neighboring subdomains for short-range interactions. LAMMPS also uses several collective communication operations: \texttt{MPI\_AllReduce} (which dominates overall communication and can lead to network congestion) aggregates data across all processors; \texttt{MPI\_Broadcast} distributes control information; and \texttt{MPI\_AlltoAll} is used during FFT computations for long-range interactions, adding to the global traffic.

\begin{figure}[!t]
    \centering
    \includegraphics[width=\textwidth]{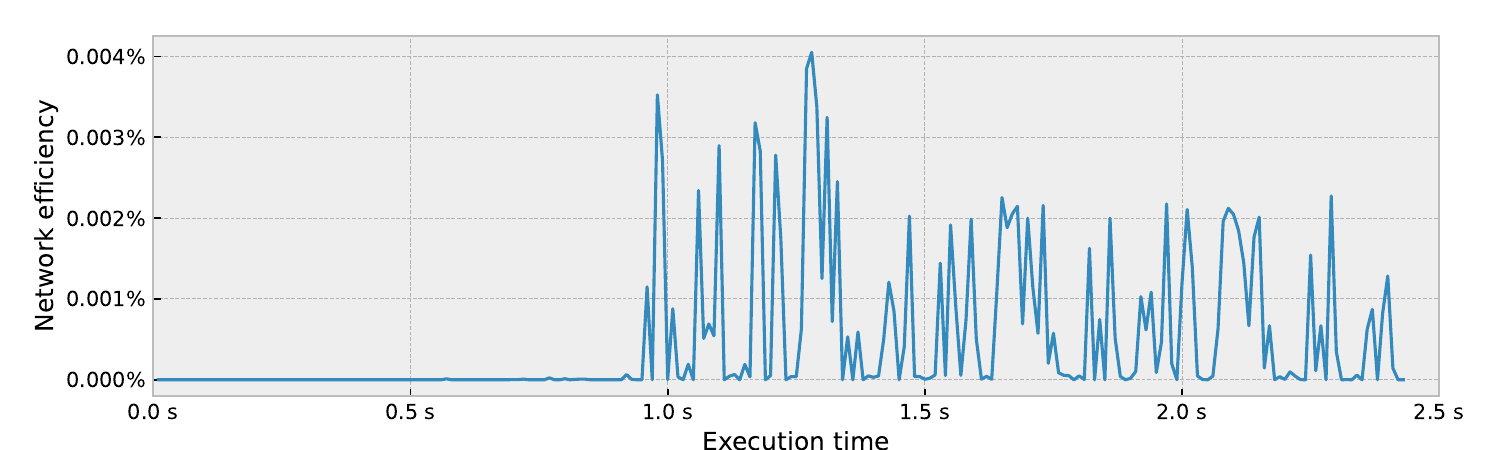}
    \caption{Network efficiency for the LAMMPS application.}
    \label{fig:lammps-throughput}
    \vspace{-.5cm}
\end{figure}

In summary, LAMMPS creates a mix of localized and global communication patterns, with \texttt{MPI\_AllReduce} identified as the primary contributor to potential network bottlenecks, especially in large-scale simulations. Being an all-to-all communication operation, \texttt{MPI\_AllReduce} will use most paths between nodes. This means that this operation will have the most degradation if all links are in a power-saving state. \figurename~\ref{fig:lammps-throughput} shows the network efficiency or throughput (normalized against the total network bandwidth) for the complete execution time (2.431 seconds) in the configured system. The total energy consumed is 9.993~MJ by nodes and 1.846~MJ by the network, totaling 11.839~MJ for the whole system. The base execution time (no power-saving mechanisms) is 2.4313 seconds.
We can see that after a second of computation, the network activity starts being intermittent with several traffic spikes due to the inter-node collective operations. 

\subsubsection{Results with fixed $t_{PDT}$ values}

\begin{figure}[!htb]
    \centering
    \adjincludegraphics[width=.45\linewidth,trim={0 {.85\height} 0 0},clip]{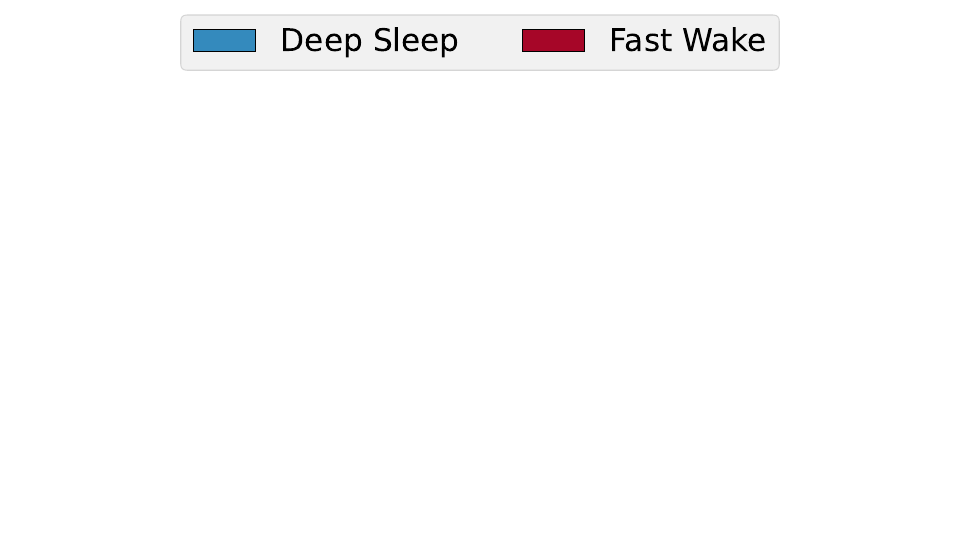}
    
    \begin{subfigure}[t]{0.49\linewidth}
        \centering
        \includegraphics[width=\linewidth]{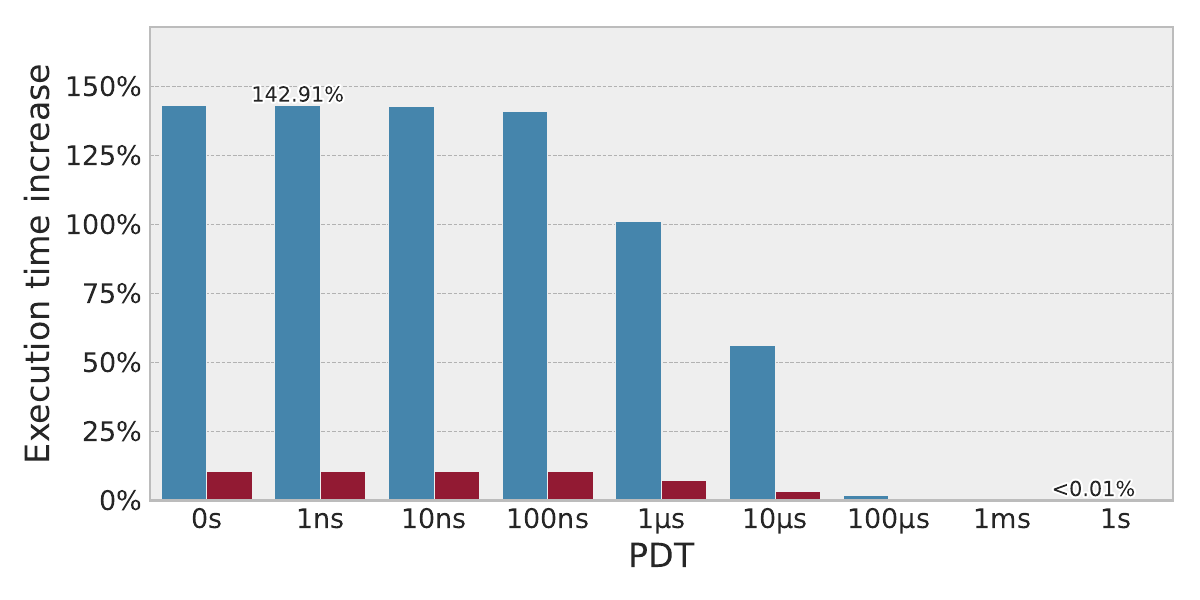}
        \caption{Execution time overhead.}
        \label{fig:LAMMPS-PDT-runtime}
    \end{subfigure}
    \begin{subfigure}[t]{0.49\linewidth}
        \centering
        \includegraphics[width=\linewidth]{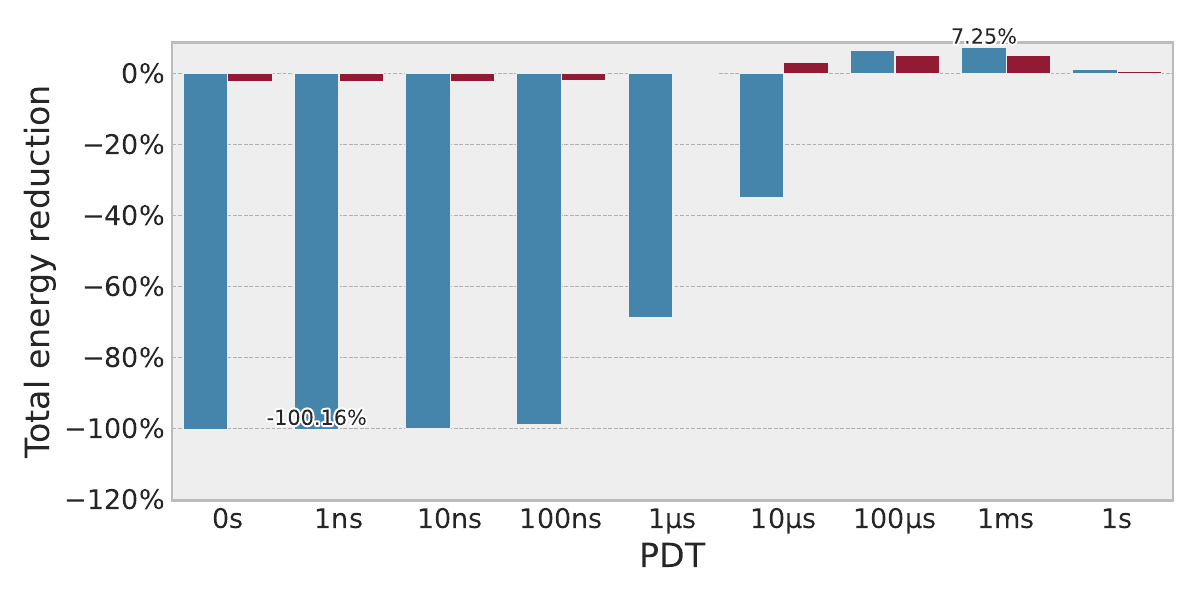}
        \caption{System energy savings.}
        \label{fig:LAMMPS-PDT-energy}
    \end{subfigure}
    
    \begin{subfigure}[t]{0.49\linewidth}
        \centering
        \includegraphics[width=\linewidth]{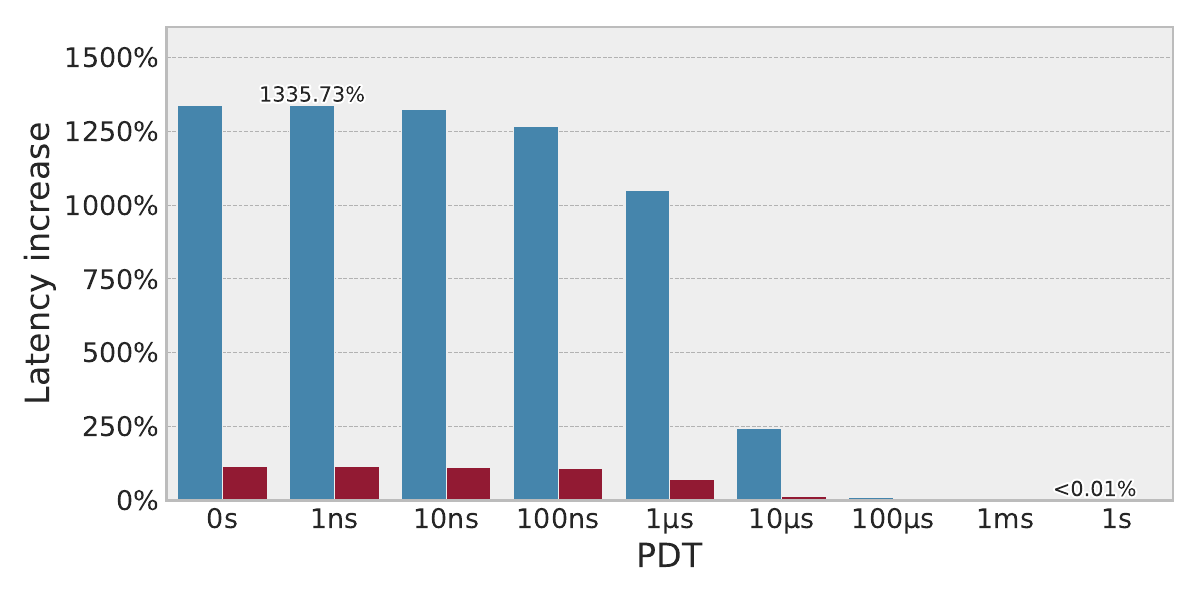}
        \caption{Packet latency increase.}
        \label{fig:LAMMPS-PDT-lat}
    \end{subfigure}
    
    \caption{Impact of different $t_{PDT}$ values for LAMMPS.}
    \label{fig:LAMMPS-PDT-all}
\end{figure}

The first statistic we will examine is the execution time, displayed in \figurename~\ref{fig:LAMMPS-PDT-runtime}. We can see that using Deep Sleep causes the execution time to go beyond 100~\%, while Fast Wake never surpasses 10~\% overhead with $t_{PDT}$ values below 10~\textmu s. This coincides with the $t_w$ for Deep Sleep being an order of magnitude above that of Fast Wake. We can see the overhead drops significantly if a $t_{PDT}$ of 100~\textmu s or higher is chosen, as this interval surpasses the length for most of the inactivity periods.

\figurename~\ref{fig:LAMMPS-PDT-all} shows the impact of different $t_{PDT}$ values on execution time, energy savings, and packet latency for LAMMPS. 
As we can see in \figurename~\ref{fig:LAMMPS-PDT-energy}, Deep Sleep achieves negative power consumption reduction when using 10~\textmu s and below, in accordance with the increment in execution time shown in \figurename~\ref{fig:LAMMPS-PDT-runtime}. 
We can also see that using Fast Wake produces an increment of power consumption with $t_{PDT}$ below 1~\textmu s, but it achieves energy savings close to 10~\% with $t_{PDT}$ values larger than 10~\textmu s. 
In this latter case, Fast Wake obtains an overhead in the execution time of less than 10~\%. 
With more permissive $t_{PDT}$ values, Deep Sleep produces more power savings.  

If we look at the packet latency overhead in \figurename~\ref{fig:LAMMPS-PDT-lat}, we can see that the tendency for each $t_{PDT}$ value is similar to that of the execution time, albeit on another order of magnitude.

In general, when using a fixed $t_{PDT}$ value, we can see that the maximum energy savings are close to 10~\% when a LAMMPS application is run using our power model and the $t_{PDT}$ values are higher than or equal to 100~\textmu s. With this $t_{PDT}$, the overheads in execution time and latency are also negligible.
Fast Wake is more sensitive to smaller $t_{PDT}$ values and achieves better energy savings with smaller $t_{PDT}$ values. Finally, note that with fixed $t_{PDT}$ values larger than 1$ms$ there are barely any energy savings.

\subsubsection{Results using PerfBound and PerfBoundCorrect}

\begin{figure}[!htb]
	\centering
	\adjincludegraphics[width=.5\linewidth,trim={0 {.85\height} 0 0},clip]{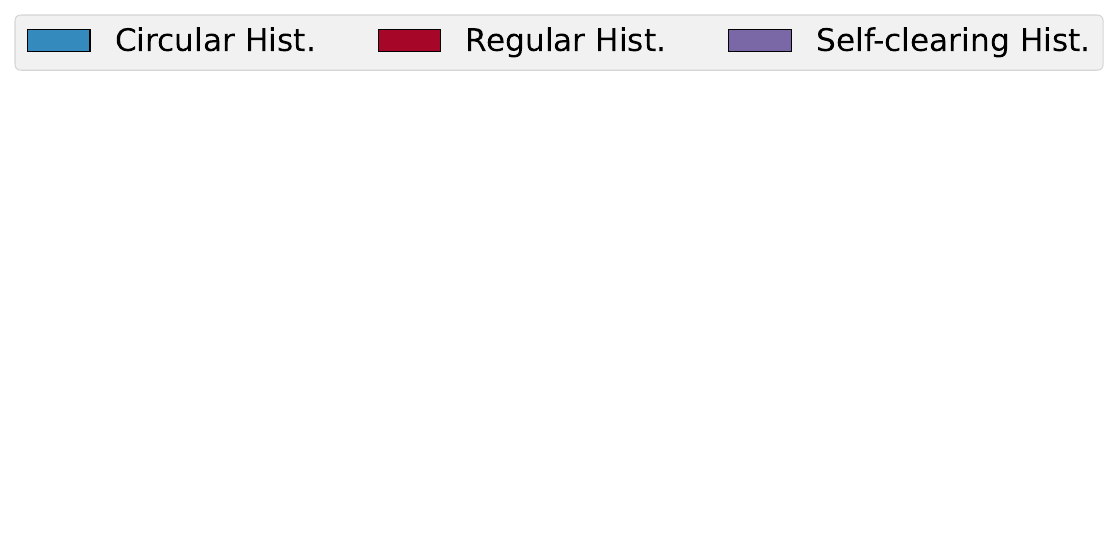}
    
	\begin{subfigure}[t]{0.49\linewidth}
		\centering
		\includegraphics[width=\linewidth]{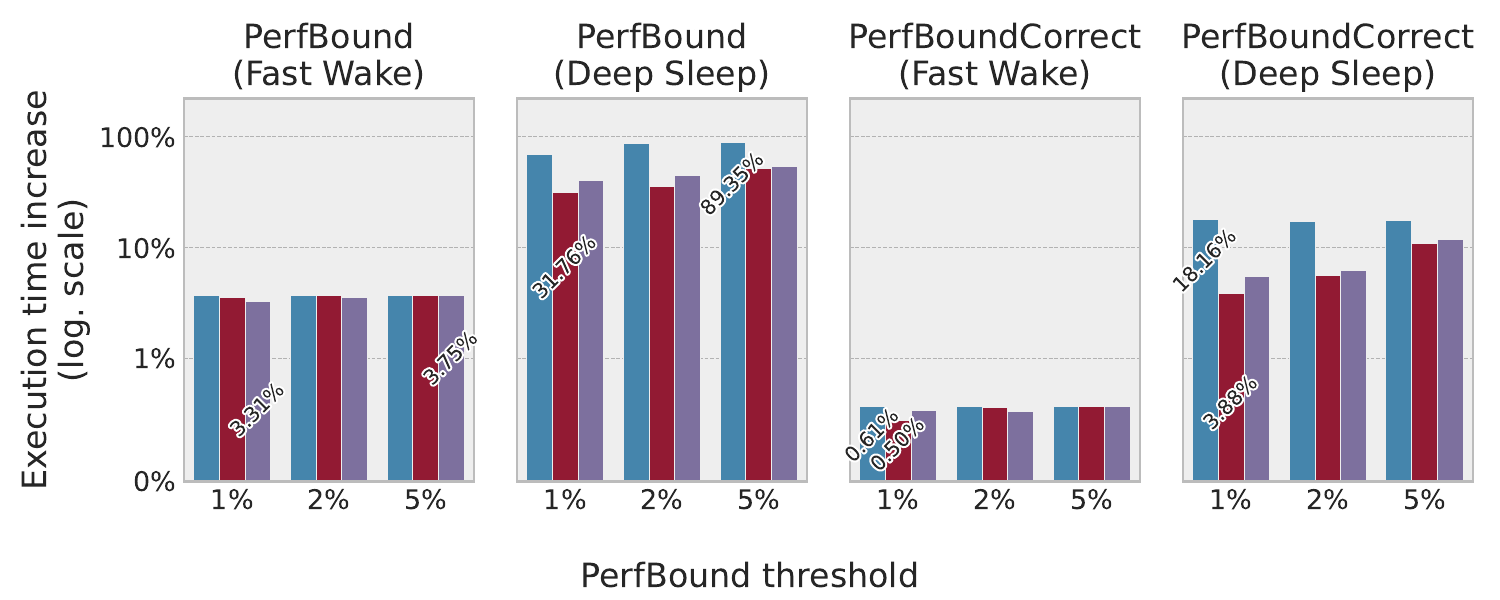}
		\caption{Execution time increase.}
		\label{fig:LAMMPS-PBOUNDvsPBOUNDC-runtime}
	\end{subfigure}
	\begin{subfigure}[t]{0.49\linewidth}
		\centering
		\includegraphics[width=\linewidth]{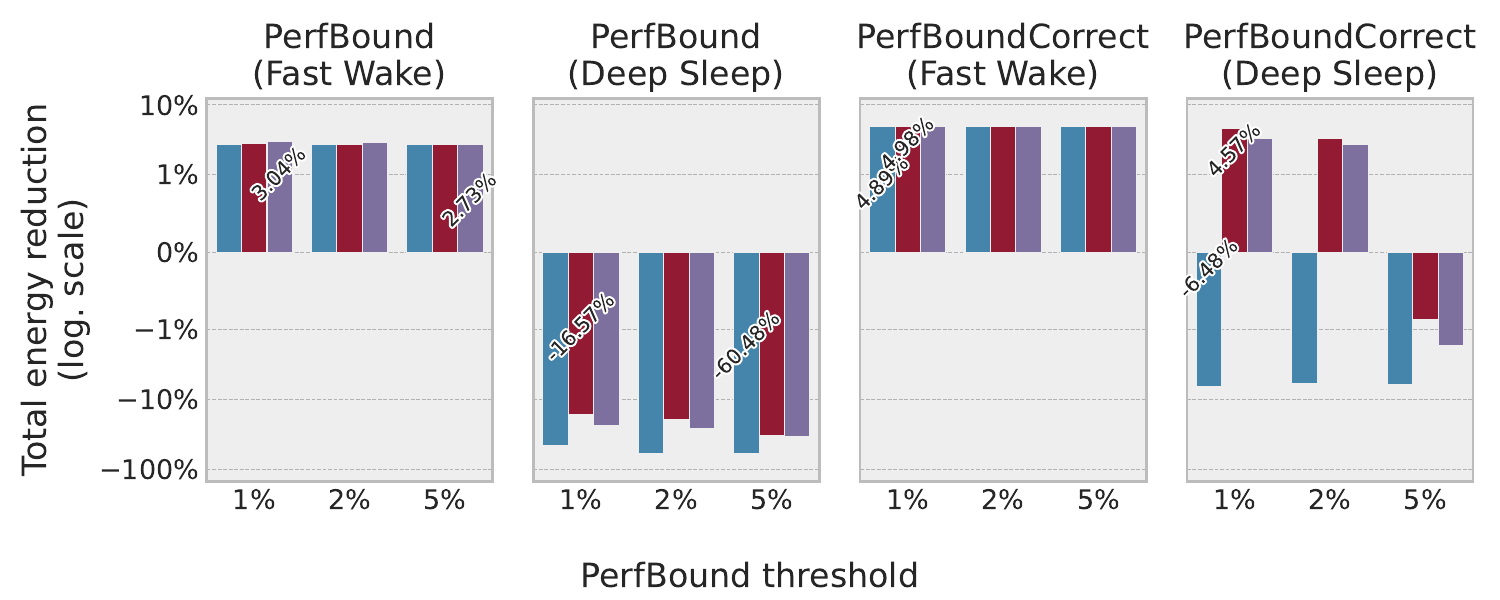}
		\caption{Energy savings.}
		\label{fig:LAMMPS-PBOUNDvsPBOUNDC-energy}
	\end{subfigure}
	
	\begin{subfigure}[t]{0.49\linewidth}
		\centering
		\includegraphics[width=\linewidth]{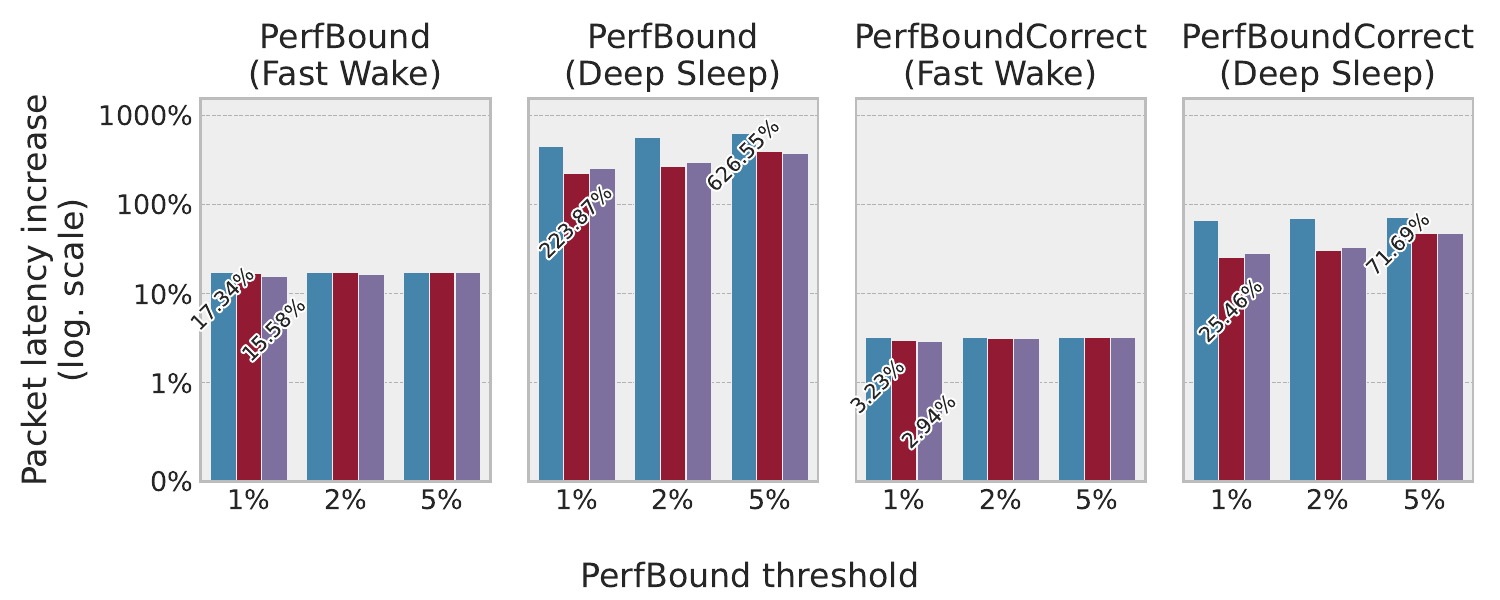}
		\caption{Packet latency increase.}
		\label{fig:LAMMPS-PBOUNDvsPBOUNDC-lat}
	\end{subfigure}
	
	\caption{Impact of PerfBound and PerfBoundCorrect on the LAMMPS trace.}
	\label{fig:LAMMPS-PBOUNDvsPBOUNDC-all}
\end{figure}

As shown in \figurename~\ref{fig:LAMMPS-PBOUNDvsPBOUNDC-runtime}, we can see the overhead on the application when using PerfBound and PerfBoundCorrect techniques. 
Just like with a fixed $t_{PDT}$, the difference between Fast Wake and Deep Sleep remains in the same proportion. 
It is important to note that using either of the histogram types makes a negligible difference in the case of Fast Wake, but has an effect when using Deep Sleep. 
Using a circular buffer to store the inactivity periods provides the worst results by a noticeable margin. 
We can also see that clearing the histogram instead of replacing old records is better for the execution time, likely because values become more permissive each time the histogram is reset. 
Recording all the inactivity periods provides a lower execution time overhead. 
However, the advantage of the self-clearing histogram comes at the cost of storing an indefinite number of bins, which is not feasible.  

We must also point out that even if PerfBound thresholds affect the calculi of every $t_{PDT}$ value throughout the simulation, their impact on the end result is not as straightforward as the original work suggests. 
Firstly, the calculations of $t_{PDT}$ are heavily dependent on the previous history. 
This poses a problem when the history record is forming, and is sensitive to irregularities and fluctuations in timings. 
Moreover, we cannot assume that the whole system is being used for our application. 
Any other traffic can disturb the formation of our records and thus, future $t_{PDT}$ values if deployed.  

The overhead on execution time has a direct impact on the energy consumed, as seen in \figurename~\ref{fig:LAMMPS-PBOUNDvsPBOUNDC-energy}. 
Both PerfBound and PerfBoundCorrect provide similar levels of energy savings, with slightly higher values using PerfBoundCorrect. 
While PerfBound with Deep Sleep does not lead to energy savings, for thresholds of 1~\% and 2~\%, PerfBoundCorrect manages to revert that situation. 
When using a high threshold, however, energy saving does not occur because the calculated $t_{PDT}$ values are too short.  

Finally, \figurename~\ref{fig:LAMMPS-PBOUNDvsPBOUNDC-lat} shows that PerfBoundCorrect, in all cases and with either power-saving state, reduces latency significantly. 
It is clear that, particularly when using Deep Sleep, a circular buffer to control the histogram size produces the highest overhead.  

We must also point out that even if PerfBound thresholds affect the calculi of every $t_{PDT}$ value throughout the simulation, their impact on the end result is not as straightforward as the original work suggests. 
Firstly, the calculations of $t_{PDT}$ are heavily dependent on the previous history. 
This poses a problem when the history record is forming, and is sensitive to irregularities and fluctuations in timings. 
Moreover, we cannot assume that the whole system is being used for our application. 
Any other traffic can disturb the formation of our records and thus, future $t_{PDT}$ values if deployed.

\subsection{Evaluation for the PATMOS application}
PATMOS~\cite{PATMOS} is a Monte Carlo neutron transport code developed at the CEA (France), designed to prototype applications in nuclear safety and radiation shielding.

To facilitate statistical analysis, simulations in PATMOS are organized into batches, even though source particles are independent. For example, one million particle histories might be simulated as 1000 batches of 1000 particles each. This batching approach allows for robust estimation of confidence intervals, as the averaged batch tally values tend to follow a normal distribution.

During each simulation cycle, particles are evenly distributed among threads, with each thread responsible for simulating particle histories and updating the associated scores. By default, static scheduling is used so that the parallelization strategy is deterministic, ensuring reproducibility and simplifying debugging.

When using MPI for distributed simulations, each MPI process independently runs a full, standalone simulation. Upon completion, results are aggregated using collective MPI operations: \texttt{MPI\_AllReduce} is used to compute the global mean, followed by \texttt{MPI\_Reduce} for variance calculation.

\begin{figure}[h!]
    \centering
    \includegraphics[width=\textwidth]{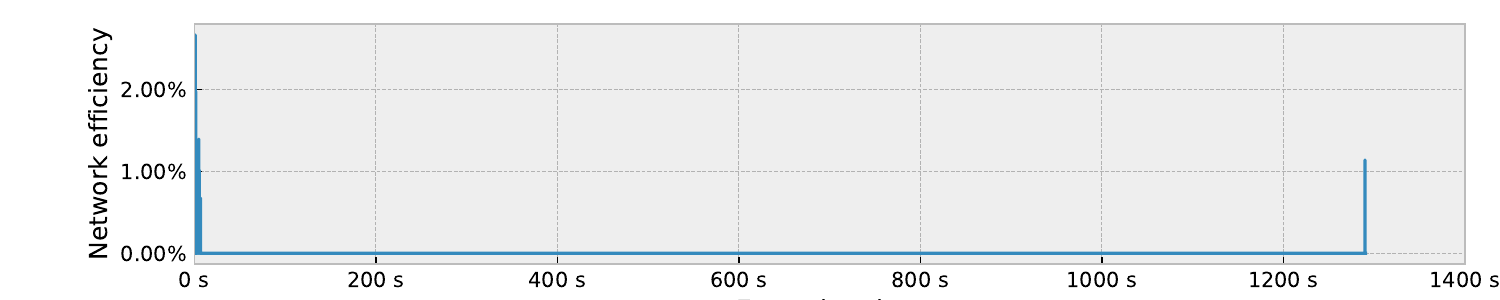}
    \caption{Network efficiency for the PATMOS application.}
    \label{fig:patmos-throughput}
    \vspace{-.5cm}
\end{figure}

This last description coincides with the evolution of network efficiency shown in \figurename~\ref{fig:patmos-throughput}. As we can see, the application only accesses the network during startup and upon completion. The rest of the time, as nodes are executing independent simulations, is an opportunity for power saving on the network.
With no power-saving strategy on links, the application execution time is 1290.64~seconds. During that time, the whole system consumed 7.334~GJ, operating at 5.685 MW, very close to maximum cluster capacity. The network has consumed 345.28~KJ. In terms of efficiency, the whole system is at 8,75E~mJ/bit, and the network is at 412,08~nJ/bit.

\subsubsection{Results with fixed ${t}_{PDT}$ values}
The first statistic we will analyze is the execution time, as it is the one that is most noticeable in terms of overhead, and longer execution times have a negative impact on total energy consumption. In our experiments, applying any power-saving technique did not affect this application. The reason behind this is that performance overhead is cumulative with the number of packets that get stalled on output ports. Therefore, as the network is only used at the start and end of the simulation, the effect on execution time is virtually non-existent.

\begin{figure}[!htb]
	\centering
	\adjincludegraphics[width=.45\linewidth,trim={0 {.85\height} 0 0},clip]{fig/plot/legend_fw-ds.pdf}
	
	\begin{subfigure}[t]{0.49\linewidth}
		\centering
		\includegraphics[width=\linewidth]{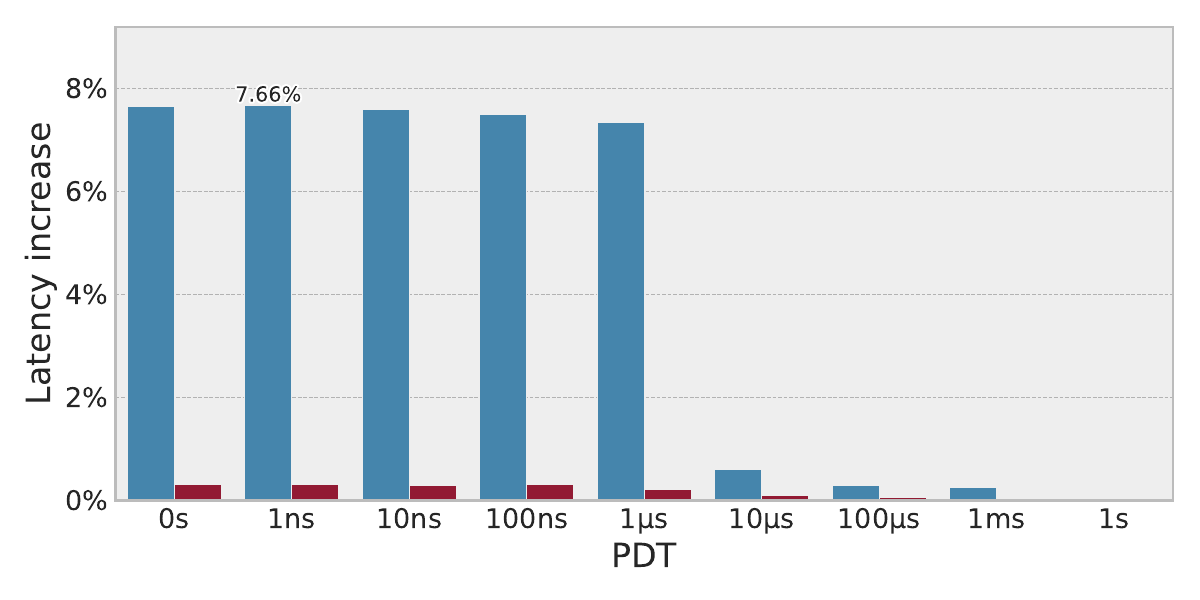}
		\caption{Packet latency increase.}
		\label{fig:PATMOS-PDT-lat}
	\end{subfigure}
	\begin{subfigure}[t]{0.49\linewidth}
		\centering
		\includegraphics[width=\linewidth]{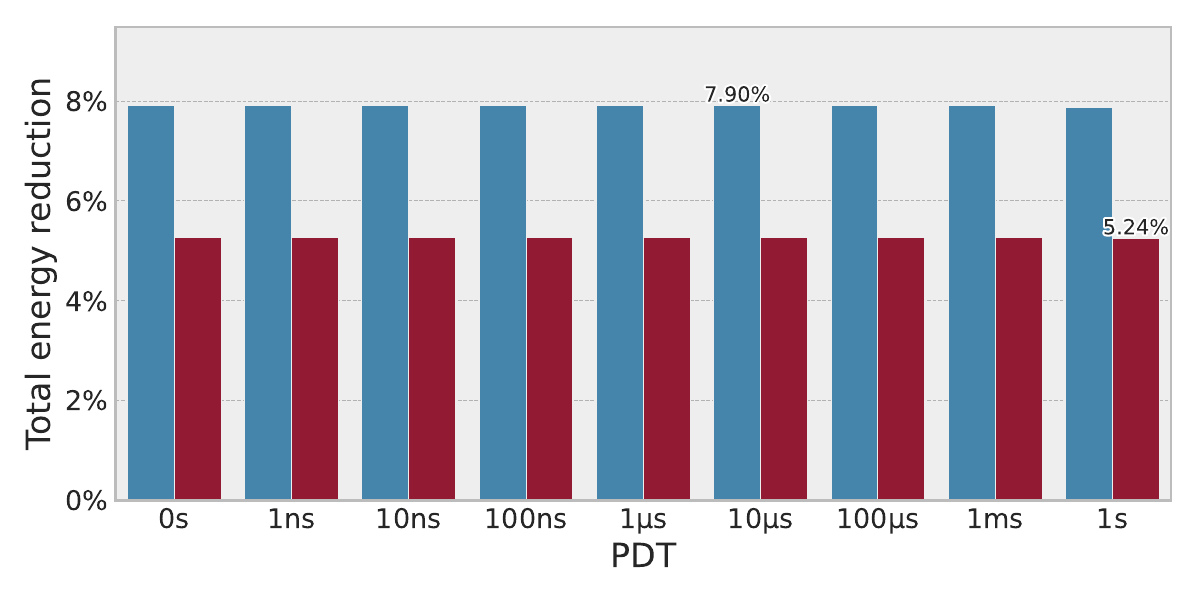}
		\caption{System energy savings.}
		\label{fig:PATMOS-PDT-energy}
	\end{subfigure}
	\caption{Impact of different $t_{PDT}$ values for PATMOS.}
	\label{fig:PATMOS-PDT-all}
\end{figure}

The other possible overhead for the trace is related to latency. If we look at \figurename~\ref{fig:PATMOS-PDT-lat}, we can see the effect each $t_{PDT}$ value has on packet latency. As we can see, employing Deep Sleep increases the packet latency by up to 7.5~\%, with a steep increase on the microsecond scale and below.

Regarding the total energy consumed, leaving links operational after transmission is potentially beneficial for future outgoing packets at the cost of more energy consumed over time. \figurename~\ref{fig:PATMOS-PDT-energy} shows the effect each PDT value has on the total energy consumed by the system. As is the case with execution time, the ports experience few transitions because of the traffic pattern of the application. As such, the energy saved on links is very high, with ports being, on average, on for only 0.02~\% of the time.

\subsubsection{Results using PerfBound and PerfBoundCorrect}
The execution time with any fixed $t_{PDT}$ value is nearly identical to the configuration without applying any power-saving. PerfBound, focusing on minimizing the overhead on execution time, is no exception. Every execution time with any threshold or histogram management scheme yields the same results.

\begin{figure}[!htb]
	\centering
	
	\adjincludegraphics[width=.5\linewidth,trim={0 {.85\height} 0 0},clip]{fig/plot/legend_hists.pdf}
	
	\begin{subfigure}[t]{0.49\linewidth}
		\centering
		\includegraphics[width=\linewidth]{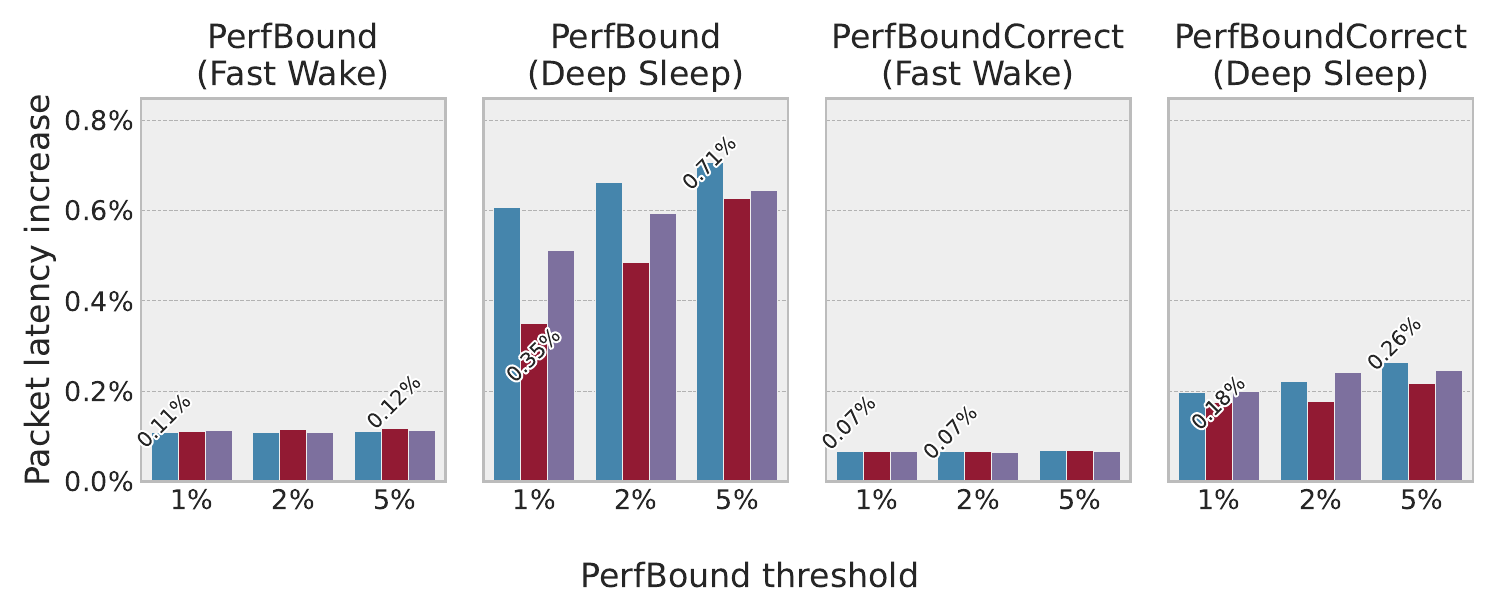}
		\caption{Packet latency increase.}
		\label{fig:PATMOS-PBOUNDvsPBOUNDC-lat}
	\end{subfigure}
	\begin{subfigure}[t]{0.49\linewidth}
		\centering
		\includegraphics[width=\linewidth]{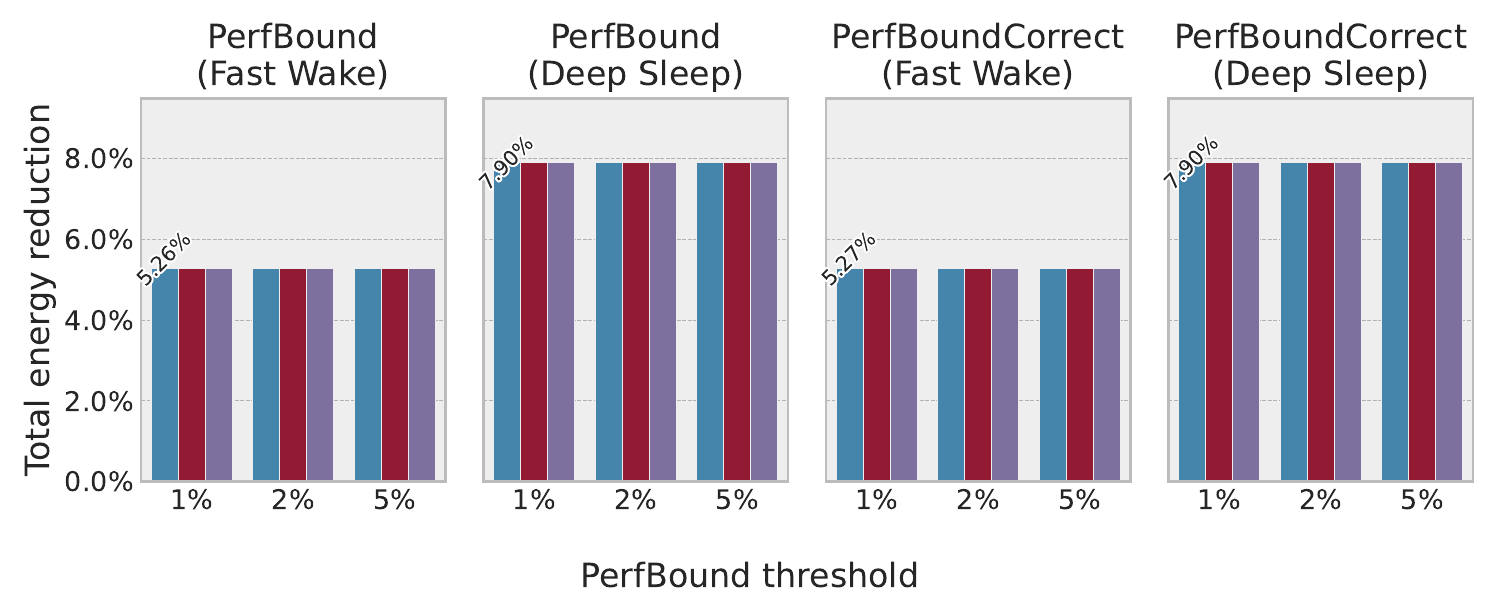}
		\caption{Energy savings.}
		\label{fig:PATMOS-PBOUNDvsPBOUNDC-energy}
	\end{subfigure}
	\caption{Impact of PerfBound and PerfBoundCorrect on the PATMOS trace.}
	\label{fig:PATMOS-PBOUNDvsPBOUNDC-all}
\end{figure}

Latency results show a minimal increase, as shown in Figure~\ref{fig:PATMOS-PBOUNDvsPBOUNDC-lat}. Deep Sleep, having a $t_w$ value an order of magnitude above that of Fast Wake, causes its latency overhead to be also higher. If we look into the different histogram strategies, we can see that having a regular histogram provides the lowest overhead of the three, no matter the PerfBound threshold chosen. This is because abnormally high values on the histogram cause the algorithm to find a suitable value early on, thus resulting in high $t_{PDT}$ values. The self-clearing histogram resets itself after 250 values collected, which will statistically remove more small values than large ones, thus resulting in more permissive $t_{PDT}$ values. And lastly, the histogram with a circular buffer, having only the latest values, has a heavy recency bias. However, all the ratios in the histogram count the same towards the geometric mean, no matter how new they are. As such, in periods where network inactivity is increasing, the predictions start to miss more often, resulting in more overhead.

When using PerfBoundCorrect, we can see that when using Deep Sleep, the latency overhead has been reduced to a third compared to PerfBound, and halved in the case of Fast Wake. We can see that the overall relation between the different histogram kinds is the same as it was with the regular PerfBound, and it is reduced in a similar proportion.

If we look at \figurename~\ref{fig:PATMOS-PBOUNDvsPBOUNDC-energy}, we can see the total energy savings on the system due to applying the power-saving techniques on the network links. We can see that for this application, there is no relevant difference in using one histogram type over the other, but using Deep Sleep yields greater energy savings. As expected in a situation with links having few and long inactivity periods, Deep Sleep provides better power saving than Fast Wake.

\subsection{Evaluation for the MLWF application}
The MLWF (Machine Learning Weather Forecast)~\cite{MLWF} application is developed within the context of the MAELSTROM European project. The trace was collected during a training session using Horovod~\cite{sergeev2018horovod2}. The traffic pattern in this application, in particular, and in Deep Learning training sessions in general, usually involves the usage of MPI collective operations for data exchange. Particularly, the most predominant traffic portion comes from \texttt{AllReduce} operations. Usually, every layer of the neural network that is being trained runs a repetitive sequence of \texttt{Gather}-\texttt{Gatherv}-\texttt{Broadcast}-\texttt{Broadcast} until the weights converge, then the \texttt{AllReduce} is invoked to have the information on every node~\cite{andujar_extending_2023}.

\begin{figure}[h!]
    \centering
    \includegraphics[width=\textwidth]{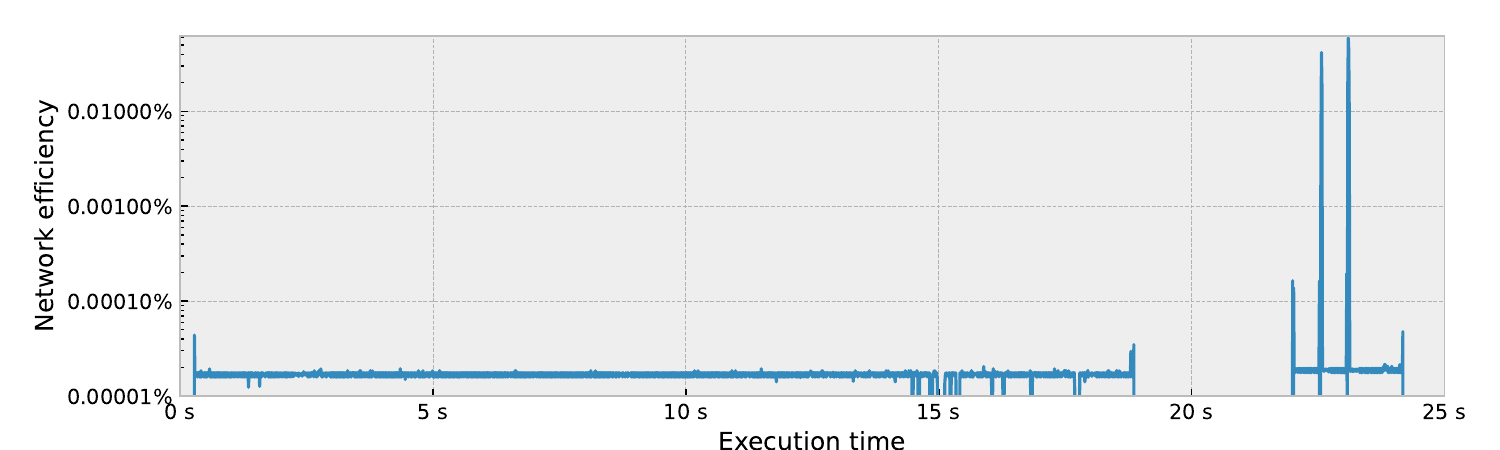}
    \caption{Network efficiency for the MLWF application.}
    \label{fig:MLWF-throughput}
    \vspace{-.5cm}
\end{figure}

The network efficiency results for this trace (\figurename~\ref{fig:MLWF-throughput}) are shown in logarithmic scale to illustrate that any traffic rate, regardless of the value, provides great difficulty in finding opportunities for power saving.

\subsubsection{Results with fixed ${t}_{PDT}$ values}

\begin{figure}[!htb]
	\centering
	
	\adjincludegraphics[width=.5\linewidth,trim={0 {.85\height} 0 0},clip]{fig/plot/legend_fw-ds.pdf}
	
	\begin{subfigure}[t]{0.49\linewidth}
		\centering
		\includegraphics[width=\linewidth]{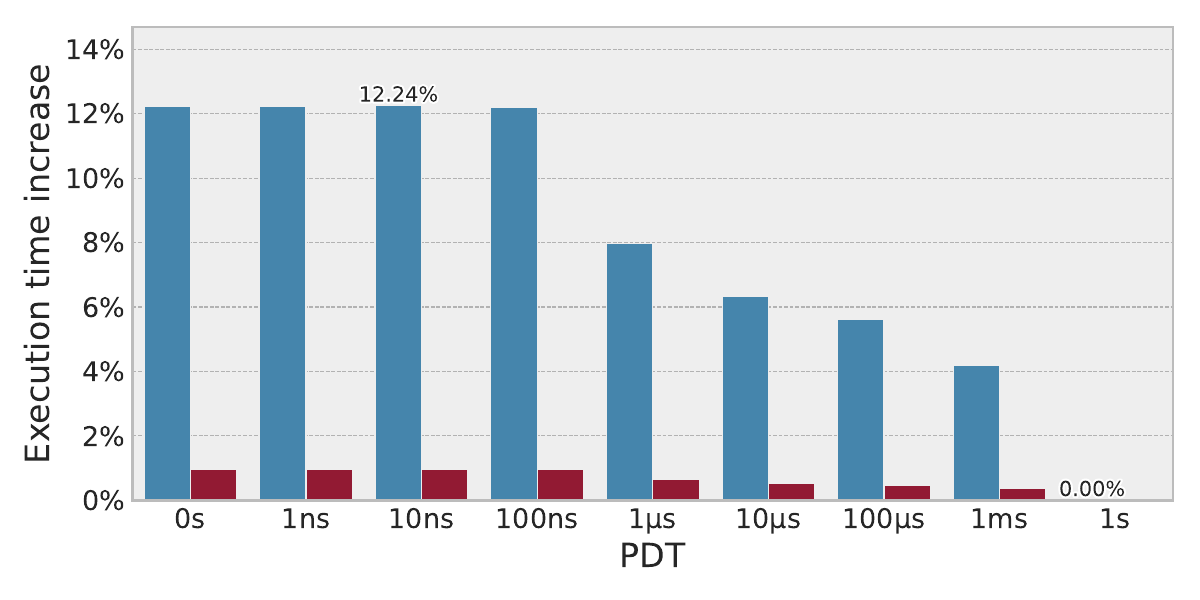}
		\caption{Execution time overhead.}
		\label{fig:MLWF-PDT-runtime}
	\end{subfigure}
	\begin{subfigure}[t]{0.49\linewidth}
		\centering
		\includegraphics[width=\linewidth]{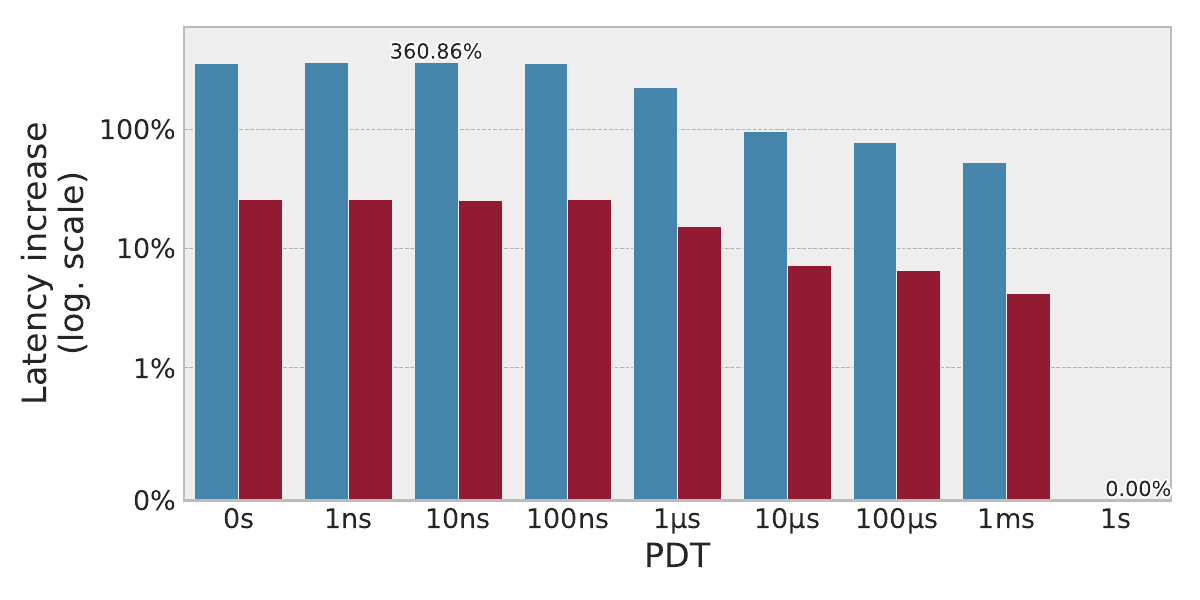}
		\caption{Packet latency increase.}
		\label{fig:MLWF-PDT-lat}
	\end{subfigure}
	
	\begin{subfigure}[t]{0.49\linewidth}
		\centering
		\includegraphics[width=\linewidth]{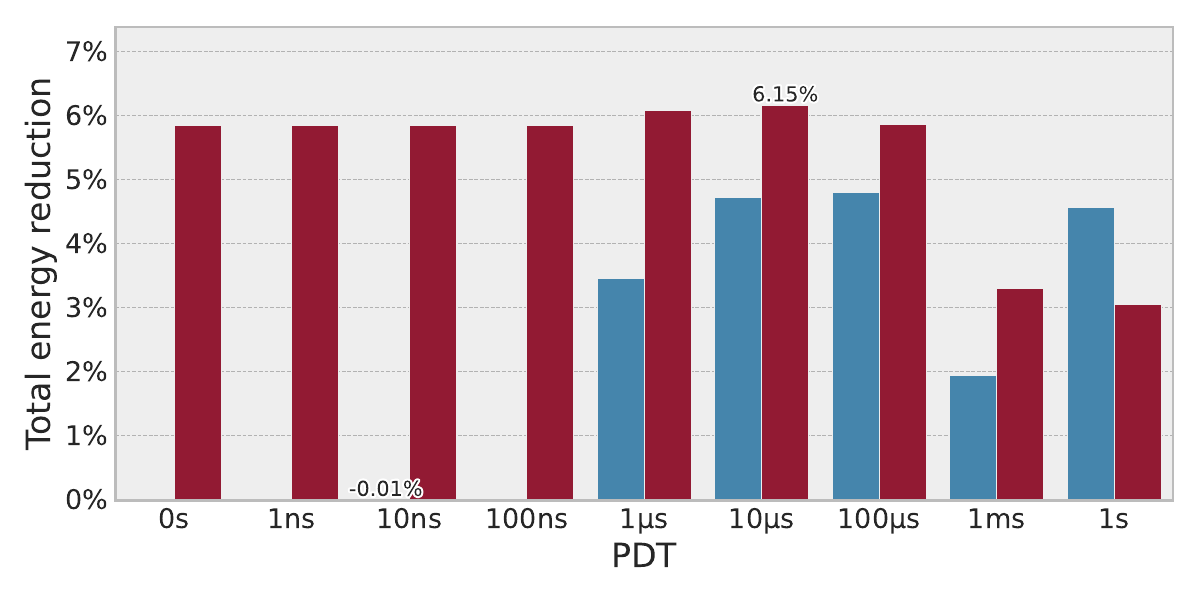}
		\caption{System energy savings.}
		\label{fig:MLWF-PDT-energy}
	\end{subfigure}
	
	\caption{Impact of different $t_{PDT}$ values for MLWF.}
	\label{fig:MLWF-PDT-all}
\end{figure}

\figurename~\ref{fig:MLWF-PDT-runtime} shows the increase in execution time for the application on different experiments. We can see that for small $t_{PST}$ values, the overhead is in consonance with the $t_w$ values for Deep Sleep and Fast Wake. To halve worst-case execution time overhead using Deep Sleep, we would have to set a $t_{PDT}$ value above 1~us, and another 3 orders of magnitude to reduce it to a third.

As shown in \figurename~\ref{fig:MLWF-PDT-lat}, we can see that there is an evident difference between the latency overhead that occurs with either power-saving state. As we increase $t_{PDT}$ values, the latency starts dropping to lower, more acceptable values.

The results for energy saving for this application are shown in \figurename~\ref{fig:MLWF-PDT-energy}. As shown, $t_{PDT}$ values of 10 and 100 \textmu s increase the energy saved in both power-saving states. Lower values provide marginal power-saving values or even increased energy usage in that proportion. Indeed, even if the traffic between the large bursts is irrelevant in terms of bytes transmitted, that communication is not negligible for power saving. However, once we use higher values, the energy reduction begins to diminish. This is due to inactivity, 99~\% of periods being within the millisecond range. This is also the reason why using a 1-second $t_{PDT}$ provides more power saving than a millisecond-scale one.

\subsubsection{Results using PerfBound and PerfBoundCorrect}

\begin{figure}[!htb]
	\centering
	
	\adjincludegraphics[width=.5\linewidth,trim={0 {.85\height} 0 0},clip]{fig/plot/legend_hists.pdf}
	
	\begin{subfigure}[t]{0.49\linewidth}
		\centering
		\includegraphics[width=\linewidth]{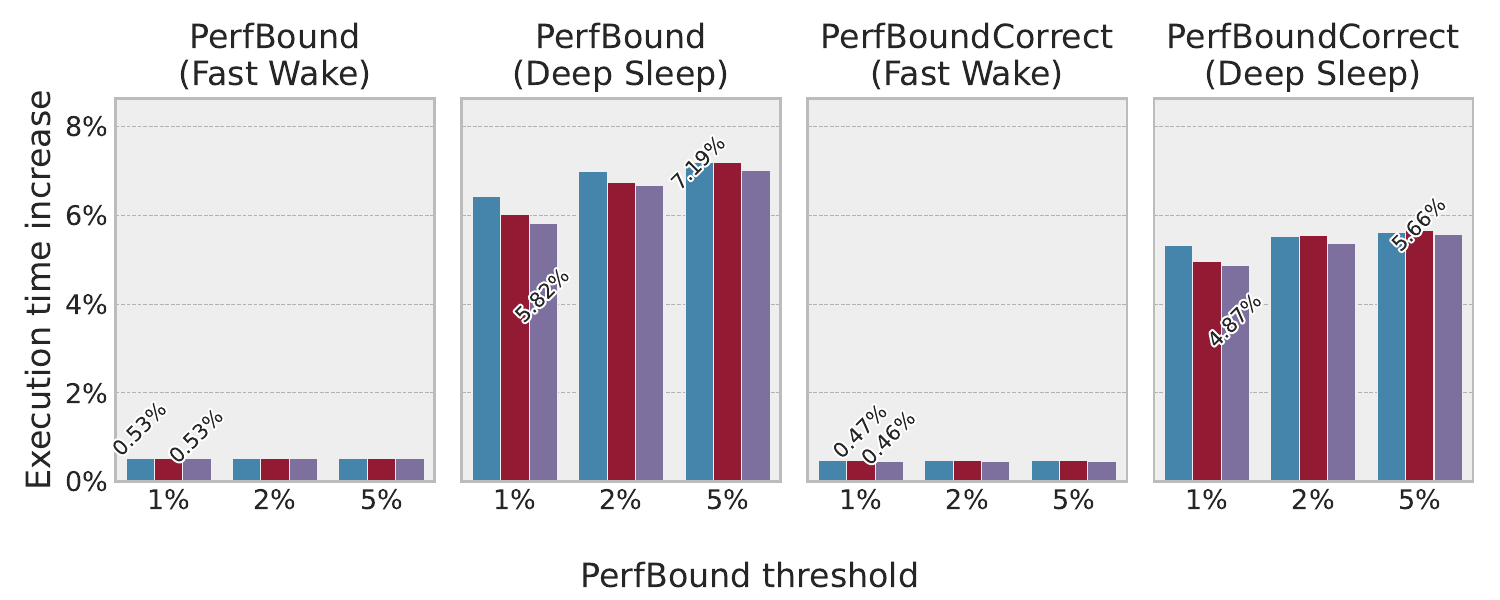}
		\caption{Execution time increase.}
		\label{fig:MLWF-PBOUNDvsPBOUNDC-runtime}
	\end{subfigure}
	\begin{subfigure}[t]{0.49\linewidth}
		\centering
		\includegraphics[width=\linewidth]{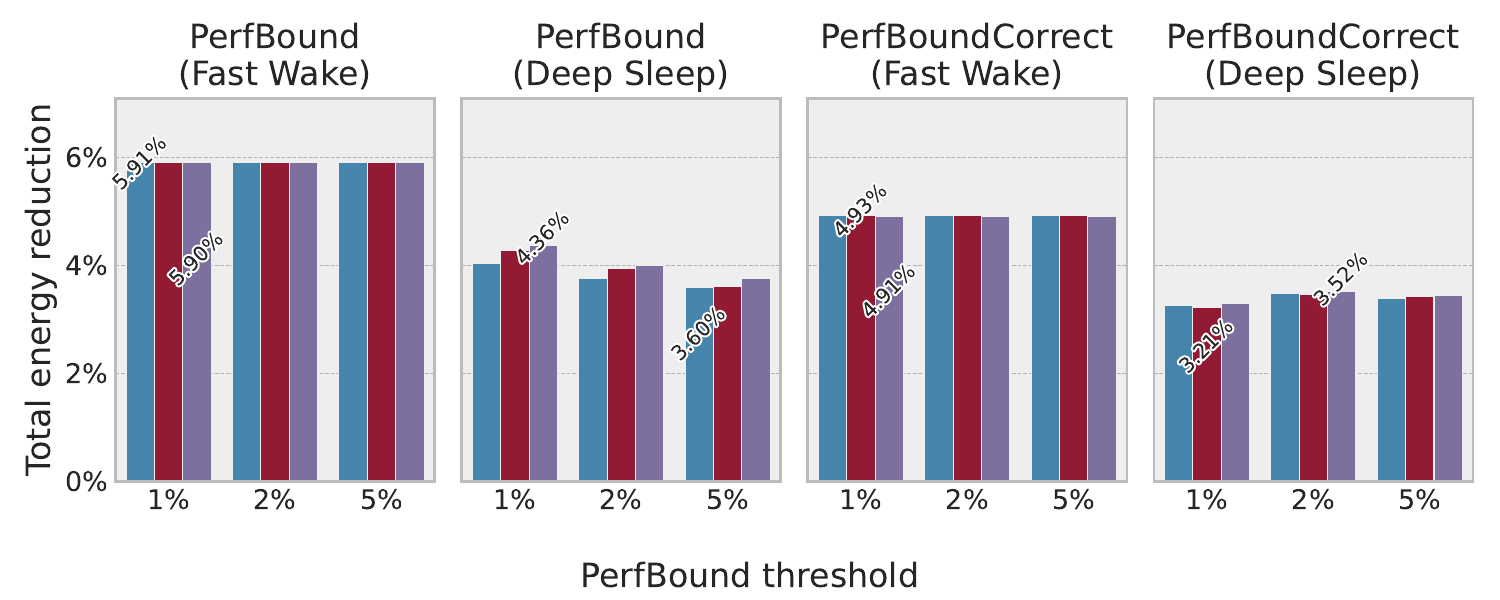}
		\caption{Energy savings.}
		\label{fig:MLWF-PBOUNDvsPBOUNDC-energy}
	\end{subfigure}
	
	\begin{subfigure}[t]{0.49\linewidth}
		\centering
		\includegraphics[width=\linewidth]{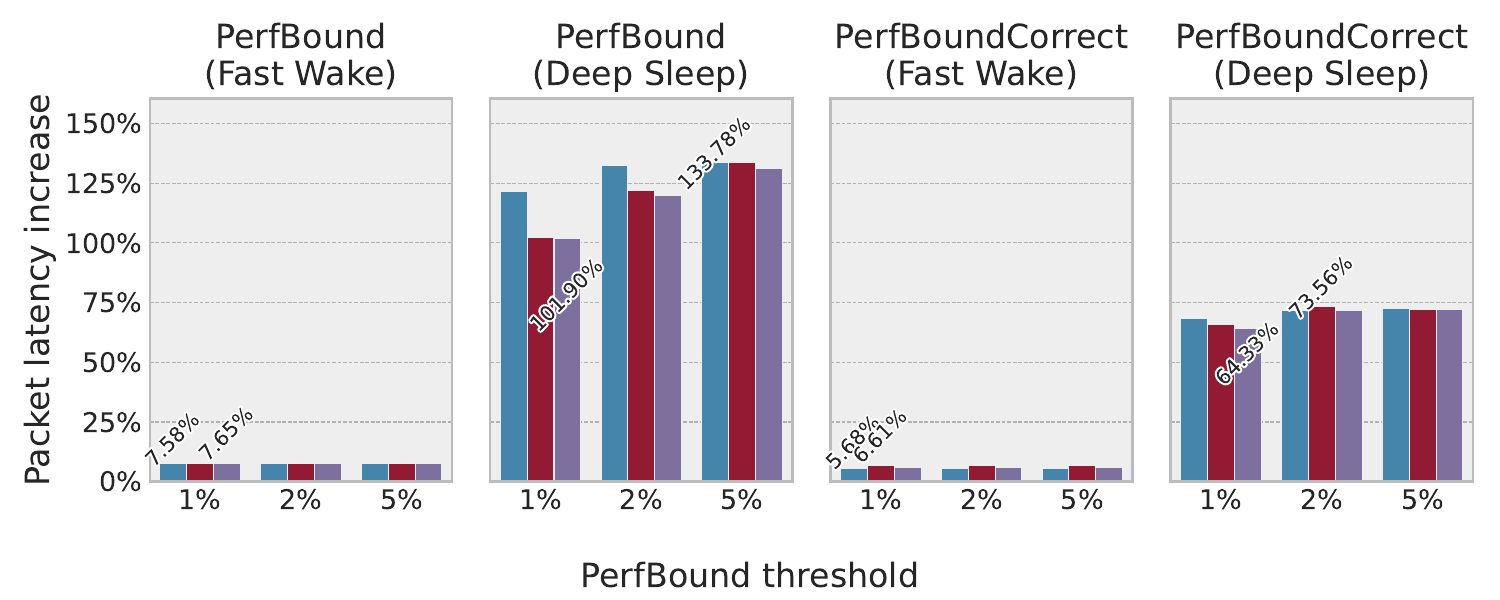}
		\caption{Packet latency increase.}
		\label{fig:MLWF-PBOUNDvsPBOUNDC-lat}
	\end{subfigure}
	
	\caption{Impact of PerfBound and PerfBoundCorrect on the MLWF trace.}
	\label{fig:MLWF-PBOUNDvsPBOUNDC-all}
\end{figure}

\figurename~\ref{fig:MLWF-PBOUNDvsPBOUNDC-runtime} shows the overhead on execution time when using PerfBound and PerfBoundCorrect with either power-saving state on links. We can see that using Fast Wake impacts the execution time by less than 1~\%, while the overhead is higher on Deep Sleep, due to its $t_w$ value. Still, we can see that PerfBoundCorrect reduces the overhead compared to PerfBound.

Regarding energy savings results displayed in \figurename~\ref{fig:MLWF-PBOUNDvsPBOUNDC-energy}, we can see that for this application, PerfBoundCorrect provides around 2~\% less energy savings than PerfBound when employing Deep Sleep, and a smaller difference on the Fast Wake experiments. The results for energy consumed show similar trends as those for execution time (\figurename~\ref{fig:MLWF-PBOUNDvsPBOUNDC-runtime}), meaning that for this application, having more permissive $t_{PDT}$ values simply increments the time that links are on. This is because PerfBound already hits most of its predictions. 

The packet latency results are shown in \figurename~\ref{fig:MLWF-PBOUNDvsPBOUNDC-lat}. We can see that for both Deep Sleep and Fast Wake, the overhead is very small. This further proves that PerfBound already makes good predictions, and PerfBoundCorrect produces diminishing returns.

\subsection{Evaluation for the AlexNet application}
\begin{figure}[h!]
    \centering
    \includegraphics[width=\textwidth]{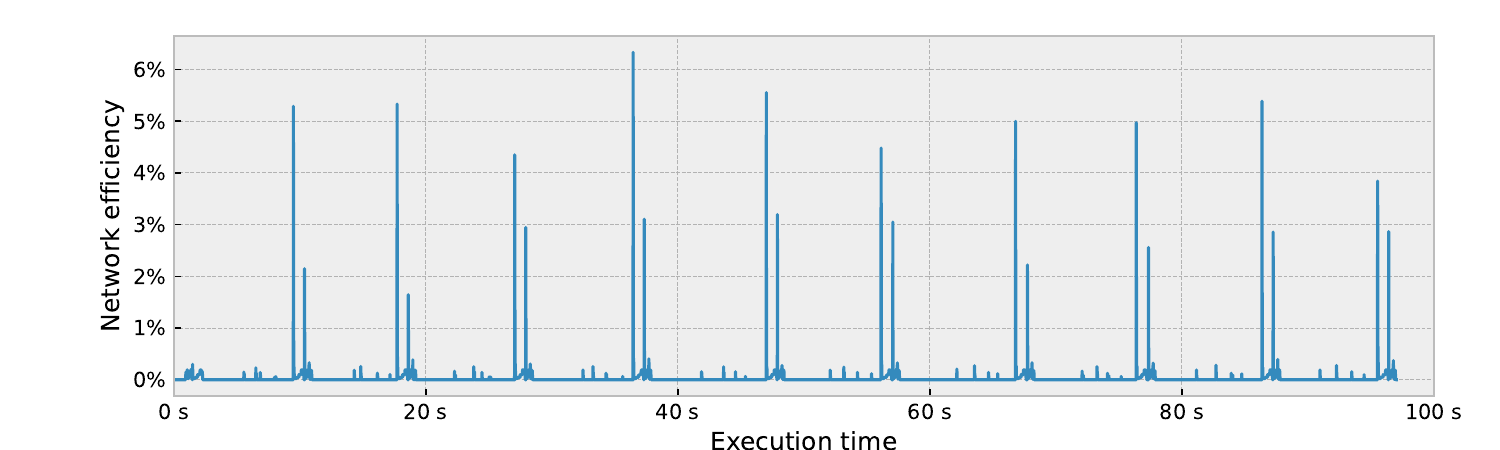}
    \caption{Network efficiency for the ALEXNET application.}
    \label{fig:ALEXNET-throughput}
\end{figure}

Our AlexNet traffic pattern is extracted from a training session, and it is shown in \figurename~\ref{fig:ALEXNET-throughput}. During this time, the main collective communications happen mainly during the back-propagation after each layer has been trained, and gradients are averaged by means of an AllReduce operation. We can see that the network is mostly idling, but there are periodic traffic bursts due to the collective communications. During the computation-heavy periods, however, we have opportunities for power-saving. Without any power-saving applied, the system consumed 397.437~MJ in total, of which 323.786~MJ (81.469~\%) are consumed by the nodes, and the rest comes from the network.

\subsubsection{Results with fixed ${t}_{PDT}$ values}

\begin{figure}[!htb]
	\centering
	
	\adjincludegraphics[width=.5\linewidth,trim={0 {.85\height} 0 0},clip]{fig/plot/legend_fw-ds.pdf}
	
	\begin{subfigure}[t]{0.49\linewidth}
		\centering
		\includegraphics[width=\linewidth]{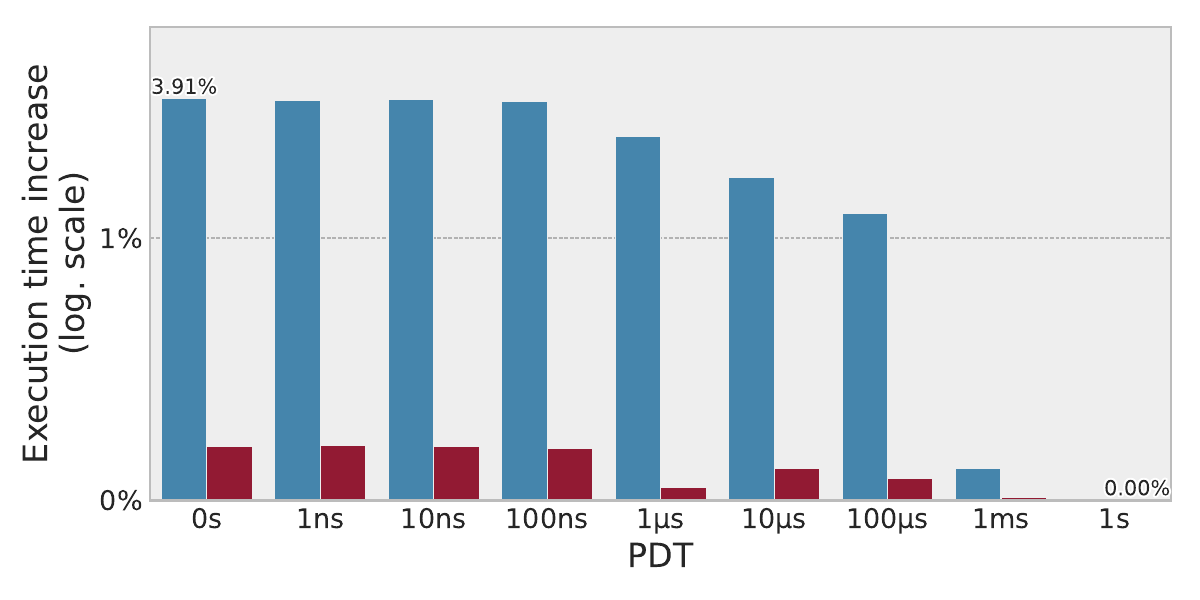}
		\caption{Execution time overhead.}
		\label{fig:ALEXNET-PDT-runtime}
	\end{subfigure}
	\begin{subfigure}[t]{0.49\linewidth}
		\centering
		\includegraphics[width=\linewidth]{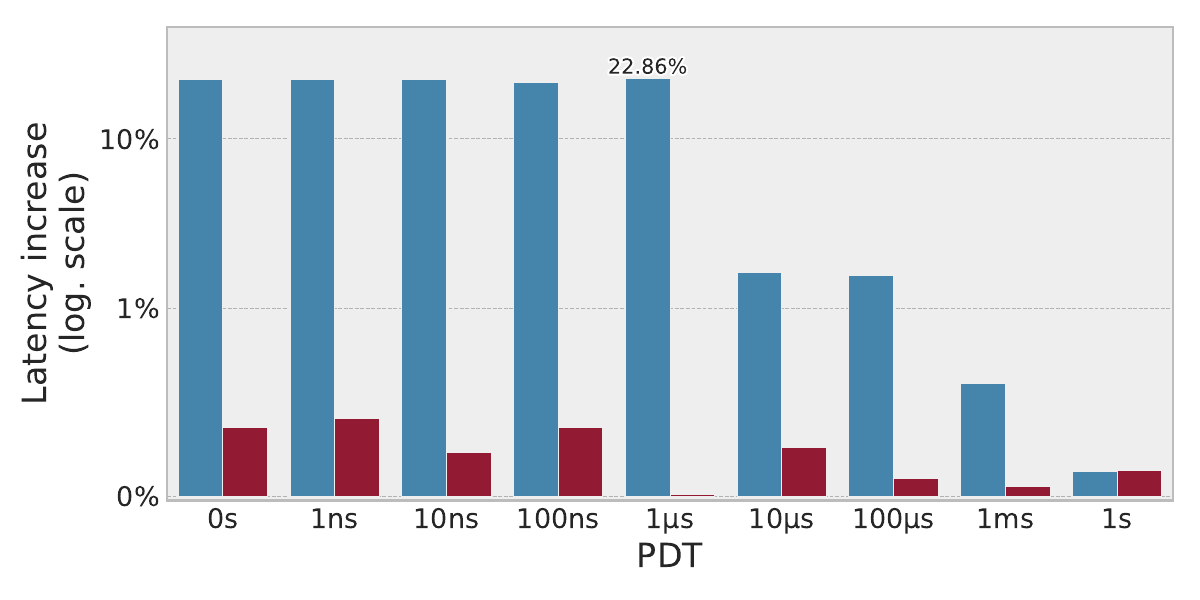}
		\caption{Packet latency increase.}
		\label{fig:ALEXNET-PDT-lat}
	\end{subfigure}
	
	\begin{subfigure}[t]{0.49\linewidth}
		\centering
		\includegraphics[width=\linewidth]{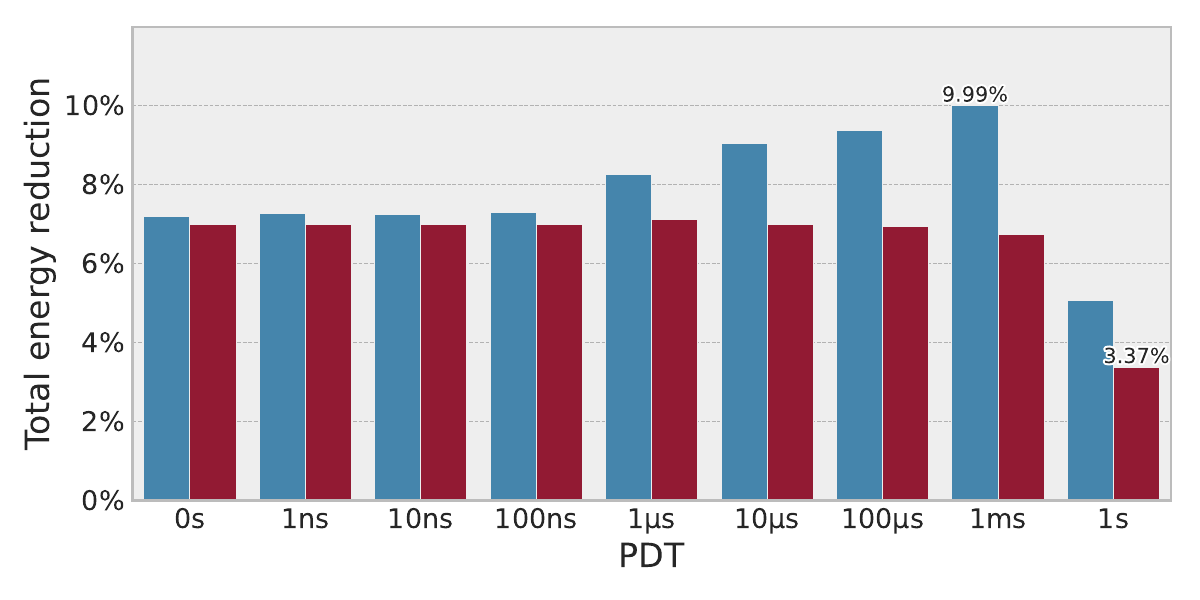}
		\caption{System energy savings.}
		\label{fig:ALEXNET-PDT-energy}
	\end{subfigure}
	
	\caption{Impact of different $t_{PDT}$ values for ALEXNET.}
	\label{fig:ALEXNET-PDT-all}
\end{figure}

\figurename~\ref{fig:ALEXNET-PDT-runtime} shows the effect on execution time caused by employing every $t_{PDT}$ value. As we can see, the overhead caused is similar to values above the nanosecond range, in which 90~\% of inactivity periods fit, as shown in \figurename~\ref{fig:inac-cdf-ALEXNET}. Because that is also the $t_w$ for Fast Wake, we find packet coalescence on output buffers on their way to their destinations, compensating for the overhead. This is why we do not see a steep increase in execution time when very restrictive $t_{PDT}$ values are employed.

Another hint of packet coalescence occurring when using Fast Wake can be seen in \figurename~\ref{fig:ALEXNET-PDT-lat}. We can see that the latency increase is minimal compared to Deep Sleep.

The benefit of using Deep Sleep instead of Fast Wake is the tradeoff of energy savings for performance. As we can see in \figurename~\ref{fig:ALEXNET-PDT-energy}, using Deep Sleep leads to more energy savings. The increase in energy saving as the $t_{PDT}$ increases is indicative that the potential for energy saving is hindered by the high number of transitions that occur. This also explains why using a $t_{PDT}$ of 1 second reverts this trend, because then it is unused ports with active PDT timers that prevent energy saving.

\subsubsection{Results using PerfBound and PerfBoundCorrect}

\begin{figure}[!htb]
	\centering
	
	\adjincludegraphics[width=.5\linewidth,trim={0 {.85\height} 0 0},clip]{fig/plot/legend_hists.pdf}
	
	\begin{subfigure}[t]{0.49\linewidth}
		\centering
		\includegraphics[width=\linewidth]{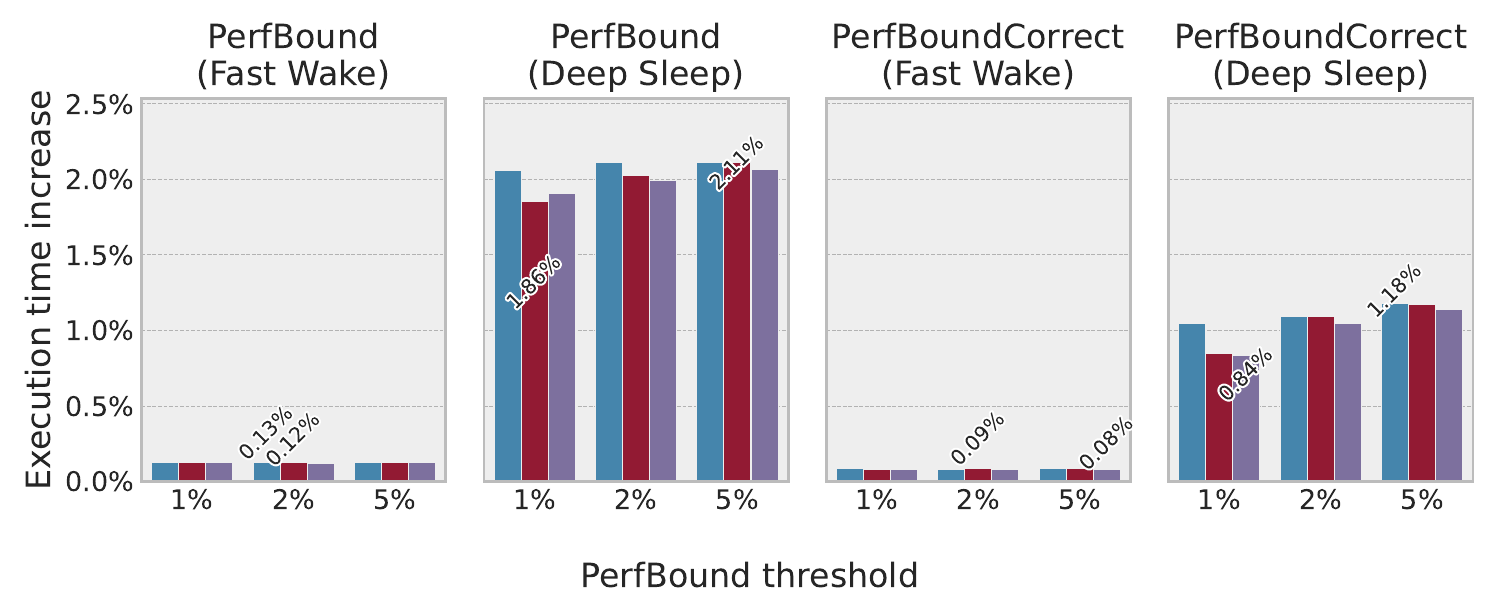}
		\caption{Execution time increase.}
		\label{fig:ALEXNET-PBOUNDvsPBOUNDC-runtime}
	\end{subfigure}
	\begin{subfigure}[t]{0.49\linewidth}
		\centering
		\includegraphics[width=\linewidth]{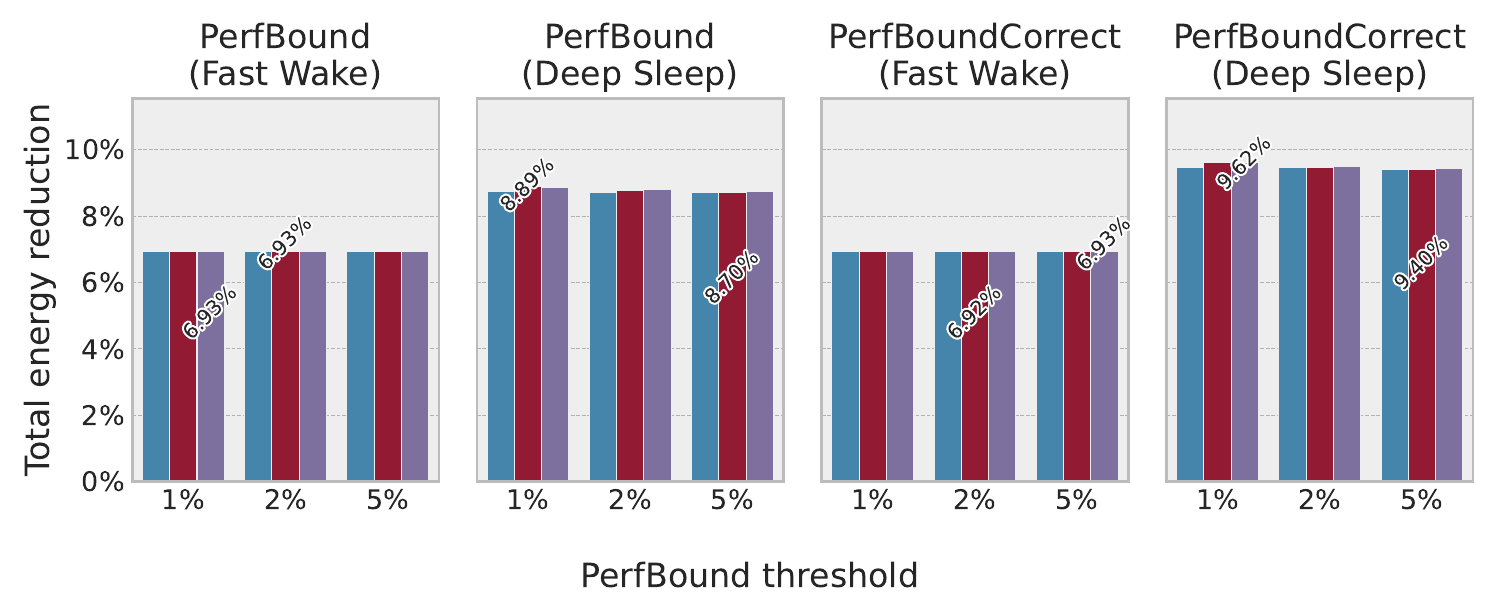}
		\caption{Energy savings.}
		\label{fig:ALEXNET-PBOUNDvsPBOUNDC-energy}
	\end{subfigure}
	
	\begin{subfigure}[t]{0.49\linewidth}
		\centering
		\includegraphics[width=.49\textwidth]{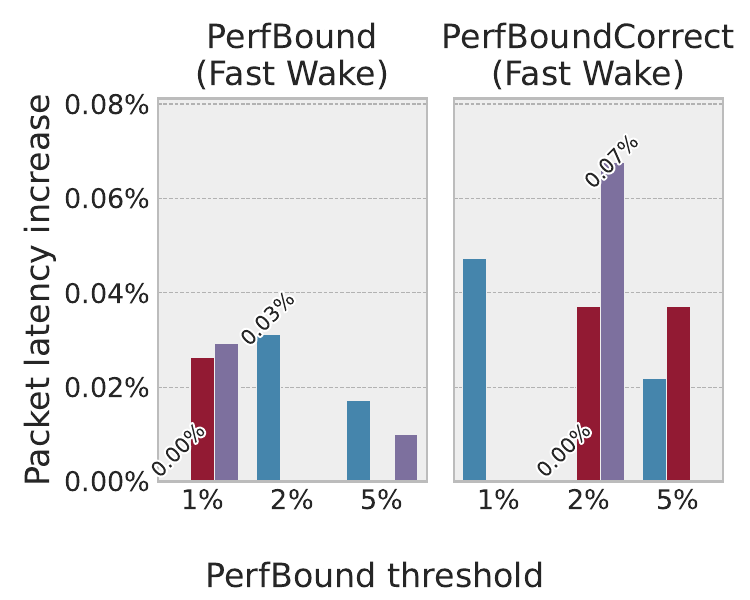}
		\includegraphics[width=.49\textwidth]{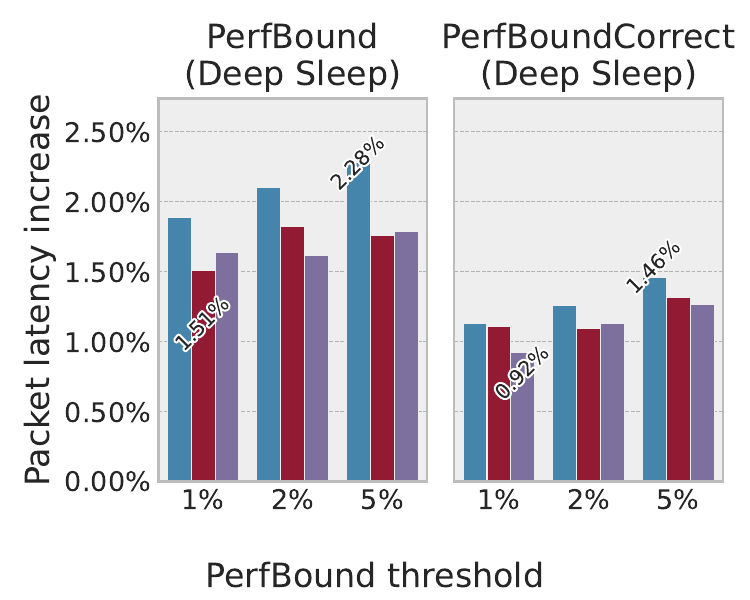}
		\caption{Packet latency increase when using PerfBound on the ALEXNET trace.}
		\label{fig:ALEXNET-PBOUNDvsPBOUNDC-lat}
	\end{subfigure}
	\caption{Impact of PerfBound and PerfBoundCorrect on the ALEXNET trace.}
	\label{fig:ALEXNET-PBOUNDvsPBOUNDC-all}
\end{figure}

As we can see in \figurename~\ref{fig:ALEXNET-PBOUNDvsPBOUNDC-runtime}, the results show the same tendencies as with the previous applications. However, this time we manage not only the same reduction in execution time overhead, but we also get that overhead to be below the performance degradation threshold using PerfBoundCorrect. 

Regarding energy savings, we can see that PerfBoundCorrect provides the same power savings as PerfBound with Fast Wake and a slight improvement using Deep Sleep.

For packet latency, the results are shown in different figures for Fast Wake and Deep Sleep due to the difference in the obtained results. We can see in \figurename~\ref{fig:ALEXNET-PBOUNDvsPBOUNDC-lat} that even if latency values are worse for PerfBoundCorrect than for PerfBound, the increase is still negligible when the packet latency is within the microsecond scale. Using Deep Sleep, even if the increase is small, we still manage to improve upon PerfBound in all cases.

\section{Outcomes and future work}
\label{sec:conclusions}

In this work, we have provided a detailed examination of the implications of power in interconnection networks. First, we explored the simplest technique by using fixed $t_{PDT}$ values, which gave us an upper bound for degradation. Later on, we discussed more sophisticated approaches and solutions that have been proposed, focusing on HPC environments using Ethernet-based networks. Indeed, BXIv3 is only one of several technologies that our proposal is applicable to, such as the upcoming OmniPath versions. After careful examination of our own implementation of the PerfBound technique and the results we gathered, we have found several key aspects that could be improved. As we have mentioned in earlier sections, the results are significant compared to the original PerfBound technique, and prove that giving more breathing room can actually be beneficial to energy efficiency. Furthermore, we also showed that other ostensibly trivial aspects, like histogram management, can have an impact on the energy saved, which was not examined in the original work.

Our observations and experiments have led us to devise a solution to mitigate the implications of using prediction for inactivity periods. Particularly, addressing avoidable prediction misses has proven useful to both performance and energy saving. Indeed, our proposal consistently outperforms PerfBound in terms of performance, while incurring minimal power overhead—or in some cases, even achieving greater energy savings.

Specifically, we have significantly reduced the overhead in terms of execution time and packet latency on the selected applications with minimal sacrifices on energy saving. In the case of the LAMMPS application, we even managed to turn an increment in energy consumption into a reduction.

In our humble opinion, our solution is subject to improvements, and several new questions arise. For example, whether providing our dynamic $t_{PDT}$ value calculation with some recency bias would improve the performance on very repetitive patterns.  

\paragraph{Acknowledgements}
This work has been supported by the Junta de Comunidades de Castilla-La Mancha under projects SBPLY/21/180225/000103, TED2021-130233B-C31 and SBPLY/21/180501/000248, and by the Spanish Ministry of Science and Innovation (MCIN) / Agencia Estatal de Investigación (AEI) under project PID2021-123627OB-C52, co-financed by the European Regional Development Fund (FEDER). We also acknowledge support from the PERTE-Chip grants (UCLM Chair, TSI-069100-2023-0014) from the Spanish Ministry of Digital Transformation and Public Service.

\bibliographystyle{unsrt}  
\bibliography{ref}  
\end{document}